\documentclass[useAMS,usenatbib,usegraphicx]{mn2e}
\usepackage{aasmacros}
\usepackage{amsmath}

\title[Properties of WDM haloes]{The properties of warm dark matter haloes}
\author[M. R. Lovell et al.]{Mark R. Lovell$^{1}$\thanks{E-mail:
    m.r.lovell@durham.ac.uk},  Carlos
  S. Frenk$^{1}$, Vincent R. Eke$^{1}$, Adrian Jenkins$^{1}$, Liang
  Gao$^{2,1}$, \newauthor and Tom  Theuns$^{1,3}$ \\ 
$^{1}$Institute for Computational Cosmology, Durham University, South
Road, Durham, UK, DH1 3LE\\ 
$^{2}$National Astronomical Observatories, Chinese Academy of Science,
  Beijing, 100012, China\\
$^{3}$Department of Physics, University of Antwerp, Groenenborgerlaan
  171, B-2020 Antwerpen, Belgium}

\newcommand{\Msun}{\mathrm{M_{\sun}}}

\newcommand{\kms}{$\rmn{km\,s^{-1}}$} 
\newcommand{\vmax}{$V_{\rmn{max}}$~}

\newcommand{\scut}{$s_{\rmn{cut}}$~}
\def\gsim{ \lower .75ex \hbox{$\sim$} \llap{\raise .27ex \hbox{$>$}} }
\def\lsim{ \lower .75ex \hbox{$\sim$} \llap{\raise .27ex \hbox{$<$}} }

\begin{document}

  \date{Accepted *** Received ***; in original
    form ***} 

  \pagerange{\pageref{firstpage}--\pageref{lastpage}} \pubyear{2012}

  \maketitle

  \label{first page}

  \begin{abstract}
    Well-motivated elementary particle candidates for the dark matter,
    such as the sterile neutrino, behave as warm dark matter (WDM).
    For particle masses of order a keV, free streaming produces a
    cutoff in the linear fluctuation power spectrum at a scale
    corresponding to dwarf galaxies. We investigate the abundance and
    structure of WDM haloes and subhaloes on these scales using high
    resolution cosmological N-body simulations of galactic haloes of
    mass similar to the Milky Way's. On scales larger than the
    free-streaming cutoff, the initial conditions have the same power
    spectrum and phases as one of the cold dark matter (CDM) haloes
    previously simulated by Springel et al. as part of the Virgo
    consortium Aquarius project.  We have simulated four haloes with
    WDM particle masses in the range $1.5-2.3$~keV and, for one case,
    we have carried out further simulations at varying resolution.
    N-body simulations in which the power spectrum cutoff is resolved
    are known to undergo artificial fragmentation in filaments
    producing spurious clumps which, for small masses ($<10^{7}\Msun$
    in our case) outnumber genuine haloes.  We have developed a robust
    algorithm to identify these spurious objects and remove them from
    our halo catalogues.  We find that the WDM subhalo mass function
    is suppressed by well over an order magnitude relative to the CDM
    case for masses $<10^{9}\Msun$.  Requiring that there should be at
    least as many subhaloes as there are observed satellites in the
    Milky Way leads to a conservative lower limit to the (thermal
    equivalent) WDM particle mass of $\sim 1.5\rmn{keV}$. WDM haloes
    and subhaloes have cuspy density distributions that are well
    described by NFW or Einasto profiles. Their central densities are
    lower for lower WDM particle masses and none of the models we have
    considered suffer from the ``too big to fail'' problem recently
    highlighted by Boylan-Kolchin et al.
  \end{abstract}


  \section{Introduction}
  \label{intro}
  
  The identity of the dark matter remains one of the central unsolved
  problems in cosmology. Various lines of evidence, for example, data
  on the cosmic microwave background radiation, indicate that the dark
  matter is made up of non-baryonic elementary particles \citep[e.g.
  ][]{Larson_11}, but exactly which kind (or kinds) of particle are
  involved is not yet known. For the past thirty years or so attention
  has focused on cold dark matter (CDM) \cite[see][for a
  review]{Frenk_White_12}, for which there are well-motivated
  candidates from particle physics, for example, the lightest
  supersymmetric particle or neutralino \citep{Ellis_84}, or the axion
  \citep{Preskill_83}. Cold dark matter particles have negligible
  thermal velocities during the era of structure formation.

  More recently, particle candidates that have appreciable thermal
  velocities at early times, and thus behave as warm, rather than
  cold, dark matter have received renewed attention. The best-known
  example is a sterile neutrino which, if it occurs as a triplet,
  could explain observed neutrino oscillation rates and baryogenesis
  \citep[e.g. ][]{Asaka_05}. This model is known as the neutrino
  minimal standard model \citep[$\nu
  \rmn{MSM}$;][]{Boyarsky09a,Boyarsky09b}; a list of alternative
  models may be found in \citet{Kusenko09}. Warm particles are
  relativistic when they decouple from the primordial plasma and
  become non-relativistic during the radiation-dominated era. This
  causes the particles to free stream out of small perturbations,
  giving rise to a cutoff in the linear matter power spectrum and an
  associated suppresion of structure formation on small scales. When
  the particles collect at the centres of dark matter haloes, their
  non-negligible thermal velocities reduce their phase-space density
  compared to the CDM case and this can result in the formation of a
  `core' in the density profile whose size varies inversely with the
  velocity dispersion of the halo \citep{Hogan_00}. However, recent
  analytical and numerical work \citep{Maccio12,Shao13, Maccio13} has
  shown that the resulting cores are astrophysically uninteresting
  being, in particular, significantly smaller than the cores claimed
  to be present in dwarf satellites of the Milky Way
  \cite[e.g.][]{Gilmore2007,deVega_10}.

  On comoving scales much larger than the free-streaming cutoff, the
  formation of structure proceeds in very similar ways whether the
  dark matter is cold or warm and so current astronomical observations
  on those scales (larger than $\sim 1$Mpc) cannot distinguish between
  these two very different types of dark matter particles. 
  Successes of the CDM paradigm, such as the remarkable agreement of
  its predictions (in a universe dominated by a constant vacuum
  energy, $\Lambda$) with observations of temperature fluctuations in
  the cosmic microwave background radiation \citep[e.g. ][]{wmap11}
  and the clustering of galaxies \citep[e.g. ][]{cole05}, carry over,
  for the most part, to a warm dark matter (WDM) model. To distinguish
  between these two types of dark matter using astrophysical
  considerations it is necessary to resort to observations on the
  scale of the Local Group.

  Over the past decade, surveys such as SDSS \citep{York_00}, PAndAS
  \citep{Ibata_07} and Pan-STARRS \citep{Kaiser10} have begun to probe
  the Local Universe in detail. A number of new dwarf spheroidal
  (dSph) satellite galaxies have been discovered around the Milky Way
  and M31 \citep[e.g.][]{Willman05b, Walsh07, Martin_09, Bell_11,
  Martin13}.  Follow-up studies of stellar kinematics have been used
  to investigate their dynamics and mass content \citep{Walker09,
  Walker10, Wolf10, Tollerud12}. These data indicate that some dSphs
  have mass-to-light ratios of around 100, and are thus systems in
  which the properties of dark matter may be most directly accesible.
  Analyses of the number and structure of dSphs should therefore
  provide strong constraints on the nature of the dark matter.

  The luminosity function of satellites in the Local Group has now
  been determined to quite faint magnitudes \citep{Koposov08,
    Tollerud08}, confirming that there are far fewer satellites around
  galaxies like the Milky Way than there are subhaloes in cosmological
  N-body simulations from CDM initial conditions
  \citep{Diemand_05,Springel_05}. This discrepancy is not new and can
  be readily explained by the physics of galaxy formation because
  feedback processes are very efficient at suppressing the formation
  of galaxies in small haloes
  \citep{Bullock_00,Benson_02,Somerville_02}. Recent hydrodynamic
  simulations have confirmed this conclusion originally deduced from
  semi-analytical models of galaxy formation \citep{Okamoto_Frenk_10,Wadepuhl_Springel_11}.

  Kinematical studies of the bright Milky Way satellites can constrain
  the internal structure of their dark matter subhaloes.
  \citet{Gilmore2007} argued that the data support the view that dSphs
  have central cores, in apparent contradiction with the results of
  N-body simulations which show that CDM haloes and their subhaloes have
  central cusps \citep{NFW_96, NFW_97, Springel_05}. \cite{Strigari10}
  explicitly showed that it is always possible to find CDM subhaloes formed in the
  Aquarius high resolution simulations of galactic haloes
  \citep{Springel08b} that are consistent with these data, however the
  subhaloes that best fit the
  kinematical data for the bright satellites turn out {\em not} to be
  the most massive ones, as would naturally be expected for these
  bright satellites. This surprising result was explored in detail in
  the Aquarius simulations by \citet{BoylanKolchin11,BoylanKolchin12},
  who dubbed it the `too big to fail' problem; it was also found in
  gasdynamic simulations of Aquarius haloes by \citet{Parry2012}.  The
  discrepancy has attracted a great deal of attention because it could
  potentially rule out the existence of CDM. Possibly
  related problems include the paucity of galaxies in voids
  \citep{Tikhonov09}, and the local HI velocity width function
  \citep{Zavala08, Papastergis2011} \citep[but see ][]{Sawala_13}.

  A number of solutions to the `too big to fail' problem have now
  been proposed.  Within the CDM context, perhaps the simplest is that
  the virial mass of the Milky Way halo is smaller than
  the average mass, $M_{200}\sim 1.4\times 10^{12}\Msun$, of the Aquarius haloes
  {\citep{VeraCiro13, Wang12}. A somewhat more uncertain possibility
    is that the central density of CDM subhaloes may have been reduced
    by the kind of explosive baryonic processes proposed by
    \cite{Navarro_97} which appear to occur in some recent
    hydrodynamic simulations \citep{Pontzen_Governato_11,
      Brooks2014,Parry2012,Zolotov2012} but not in others
    \citep{diCintio11} which assume different prescriptions
    for physics that are not resolved in the simulations. 

    More radical solutions to the `too big to fail' problem require
    abandoning CDM altogether.  \citet{Vogelsberger12}
    show that simulations with a new class of `self-interacting'
    dark matter could solve the problem. However, a solution is also
    possible with more conventional assumptions. In particular,
    \citet{Lovell12} show that simulations with WDM produce very good
    agreement with the dSph kinematical data. The absence of
    small-scale power in the initial fluctuation field causes structure to
    form later than in the CDM case. Haloes of a given mass thus
    collapse when the mean density of the universe is smaller and, as
    a result, end up with lower central densities
    \citep{AvilaReese01}. However, the WDM
    model they assumed was `too warm', in the sense that it assumed
    too low a particle mass (and thus too large a cut-off scale in the
    initial power spectrum) and produced only 18 dark matter subhaloes
    within 300~kpc of the main halo centre whereas observations suggest
    the actual number of satellites may be over an order of
    magnitude greater \citep{Tollerud08}. 

    This constraint from subhalo central densities is one of several
    that can be used to place bounds on the WDM particle
    mass.  The measured clustering of the Lyman~$\alpha$ forest lines
    at high redshift sets a lower limit to the particle mass
    \citep{Viel05,Boyarsky09b,Viel_13} while the absence of X-rays
    from particle decay sets a (model dependent) upper limit to the
    mass of the sterile neutrino (whose decay rate into pairs of
    neutrinos and X-ray photons scales with the mass of the sterile
    neutrino; see \citealt[and references
    therein.]{Kusenko09,Boyarsky12})
    
    The results of \cite{Lovell12} and related results by
    \cite{Maccio12, Shao13, Maccio13} raise the question of whether it is possible to
    find a range of WDM particle masses that lead to `warm enough'
    models that match satellite central densities but which are also
    `cold enough' to generate the observed number of satellite
    galaxies \citep{Polisensky2011, Kamada13}. In this work we examine
    both the number and structure of satellite galaxies in simulations
    as a function of the WDM particle mass. 

    The first requirement is to be able to count accurately the number
    of dark matter haloes formed in WDM cosmologies. The first
    simulations of WDM models \citep{Bode01} showed the halo mass
    function to be suppressed as expected, but also found that at
    least 90 percent of haloes, depending on the choice of power
    spectrum cutoff, formed from the fragmentation of filaments and
    had masses below the smoothing scale.  \citet{Wang07} examined
    this effect in hot dark matter (HDM) simulations (which assume a
    much larger power spectrum cutoff scale than in WDM) and showed
    that the fragmentation of filaments depends on the resolution of
    the simulation, thus concluding that most of the haloes in the
    \cite{Bode01} simulations were due to a numerical artifact. 

  In this paper we introduce a series of methods for identifying
  spurious haloes in simulations, and then use our cleaned halo sample
  to examine the distribution and structure of WDM haloes as a
  function of the power spectrum cutoff. The paper is organised as
  follows. In Section~\ref{Sims} we present our simulation set and in
  Section~\ref{RSSsec} we describe our algorithm for removing spurious
  subhaloes. We then present our results in Section~\ref{Res}, and
  draw conclusions in Section~\ref{Conc}. 

  \section{The simulations}
  \label{Sims}

   We begin by describing the details of our simulations, the
   procedure for generating initial conditions and a general
   overview.

  \begin{table*}
    \centering
    \begin{tabular}{|c|c|c|c|c|}
      \hline
      Simulation & $m_{\rmn{WDM}} [\rmn{keV}]$ &  $\alpha [ h^{-1}\rmn{Mpc}]$ &
      $M_{\rmn{th}} [\Msun]$ & $m_{\rm{WDM}}^{\nu=1.12} [\rm{keV}]$ \\
      \hline  
      CDM-W7 & -- & 0.0 & -- & -- \\
      $m_{2.3}$ & 2.322 & 0.01987 & $1.4\times 10^{9}$ & 1.770 \\
      $m_{2.0}$ & 2.001 & 0.02357 & $1.8\times 10^{9}$ & 1.555 \\
      $m_{1.6}$ & 1.637 & 0.02969 & $3.5\times 10^{9}$ & 1.265 \\
      $m_{1.5}$ & 1.456 & 0.03399 & $5.3\times 10^{9}$ & 1.106 \\

      \hline
    \end{tabular}
    \caption{Parameters of the 
      simulations. The parameter $\alpha$ determines the power
      spectrum cutoff (Eqn.~\ref{eqn:tf}); $m_{\rmn{WDM}}$ is the thermal relic mass
      corresponding to each value of $\alpha$; and $M_{\rmn{th}}$ is
      the cutoff mass scale defined using a top hat filter as
      described in the text. The final column gives the particle
	masses that, when combined with the $\nu=1.12$ transfer
      function and $m_{\rmn{WDM}}-\alpha$ relation of
	\citet{Viel05}, give the best approximation to our $\nu=1$
	transfer functions.}
    \label{Tab1}
  \end{table*}

\subsection{Simulation parameters}
  Our N-body simulation suite is based upon that of the Aquarius
  Project \citep{Springel08b}, a set of six (Aq-A through to Aq-F)
  galactic dark matter haloes simulated at varying resolution (levels
  1-5, where level~1 corresponds to the highest resolution). The
  Aquarius simulations assumed cosmological parameter values derived
  from the {\emph Wilkinson Microwave Anisotropy Probe (WMAP)} year~1 data. These have now been superseded and in
  this paper we use the cosmological parameter values derived from
  the {\emph WMAP} year~7 data \citep{wmap11}: matter density,
  $\Omega_{m}=0.272$; dark energy density, $\Omega_{\Lambda}=0.728$;
  Hubble parameter, $h=0.704$; spectral index, $n_{s}=0.967$; and
  power spectrum normalization $\sigma_{8}=0.81$.

  Our main set of simulations follows the formation of four WDM
  galactic haloes with different effective WDM particle
  masses. The initial phases in the fluctuation spectrum are identical
  to those of the original CDM Aq-A halo but the transfer function is
  that appropriate to WDM as described below. In addition, we
  resimulated the level-2 Aq-A halo using the WMAP year-7 cosmology.
  For all five haloes (one CDM and four WDM), we ran simulations at
  different resolution. Our `high resolution' suite corresponds to
  level-2 in the original Aquarius notation; it has particle mass of
  $1.55\times10^{4}\Msun$, and gravitational softening length of
  $\epsilon=68.1\rmn{pc}$. All haloes were also run at ``low
  resolution'' (level-4), with particle mass of
  $4.43\times10^{5}\Msun$ and gravitational softening of
  $\epsilon=355.1\rmn{pc}$. Finally, we ran an intermediate resolution
  version (level 3) of the warm dark matter models with the lightest
  and heaviest dark matter particles, with particle mass $5.54\times10^{4}\Msun$ and
  $\epsilon=125.0\rmn{pc}$, in order to facilitate convergence
  studies.  All haloes were simulated from $z=127$ to $z=0$ using the
  \textsc{gadget3} N-body code \citep{Springel08b}.

  To set up the initial conditions for the WDM runs we employed the
  transfer function, $T(k)$, defined as
  \begin{equation}
    P_{\rmn{WDM}}(k) = T^2(k) P_{\rmn{CDM}}(k).
  \end{equation}
  where $P(k)$ denotes the power spectrum as a function of comoving
  wavenumber $k$. We adopted the fitting formula for $T(k)$ given by
  \citet{Bode01}:

  \begin{equation}
    T(k) = (1+(\alpha k)^{2\nu})^{-5/\nu},
    \label{eqn:tf}
  \end{equation}

  \noindent
  where $\nu$ and $\alpha$ are constants. \citet{Bode01} and
  \citet{Viel05} find that $\nu$ can take values between 1 and 1.2
  depending on the fitting procedure; we adopted $\nu=1$ for
  simplicity. The position of the cutoff in the power spectrum is
  determined by the parameter $\alpha$, such that higher values of
  $\alpha$ correspond to cutoffs at larger length scales. In principle,
  the initial conditions for WDM simulations should include thermal
  velocities for the particles \citep{Colin08,Maccio12,Shao13}.  However, at
  the resolution of our simulations, the appropriate velocities would
  have a negligible effect \citep{Lovell12} and are therefore not
  included. All of our CDM and WDM initial conditions employed a
  glass-like initial particle load \citep{White94}.

  For our four WDM models we adopted values of $\alpha$ of
  $0.0199h^{-1}\rmn{Mpc}$, $0.0236h^{-1}\rmn{Mpc}$,
  $0.0297h^{-1}\rmn{Mpc}$, and $0.0340h^{-1}\rmn{Mpc}$ respectively.
  The last of these corresponds to the original WDM simulation
  presented in \citet{Lovell12} which, however, assumed the WMAP
  year-1 cosmological parameters. That model was originally chosen as
  a thermal relic approximation to the M2L25 model of
  \citet{Boyarsky09b}, the $\nu \rmn{MSM}$ parameter combination that
  has the largest effective free-streaming length that is still
  consistent with bounds from the Lyman-$\alpha$ forest \citep[but see
  also][]{Viel_13}. 

 \citet{Bode01} related $\alpha$ to a generic thermal relic warm dark
  matter particle mass, $m_{\rmn{WDM}}$, using the formula: 

    \begin{multline}
      \alpha = 
      \frac{0.05}{h\rmn{Mpc}^{-1}}
      \left(\frac{m_{\rmn{WDM}}}{1\rmn{keV}}\right)^{-1.15}
      \left(\frac{\Omega_{\rmn{WDM}}}{0.4}\right)^{0.15} \\ 
      \times\left(\frac{h}{0.65}\right)^{1.3}
      \left(\frac{g_{\rmn{WDM}}}{1.5}\right)^{-0.29},
    \end{multline}


  
  \noindent
  where $\Omega_{\rmn{WDM}}$ is the WDM contribution  to the density
  parameter; we have set the number of degrees of freedom,
  $g_{\rmn{WDM}}=1.5$. We list the thermal relic masses for each of our models
  in Table~\ref{Tab1}, and use these masses as labels for the models,
  namely $m_{2.3}$, $m_{2.0}$, $m_{1.6}$, and
  $m_{1.5}$; we denote the CDM simulation with {\emph WMAP} year-7
  parameters as CDM-W7. We also give
  the cutoff mass scale for each
  simulation, which we define as the mass within a top hat filter
  which, when convolved with the CDM power spectrum, results in a
  function that peaks at the same value of $k$ as the WDM power
  spectrum. 

  In order to compare our study to that of \cite{Viel05} and
  \cite{Viel_13} we need to take into account that the transfer
  function that we use assumes $\nu=1$ in Eqn.~\ref{eqn:tf} while
  theirs assumes $\nu=1.12$. For values of $k$ near the power spectrum
  cutoff, the transfer function for a given $m_{\rmn{WDM}}$ has a
  higher amplitude if $\nu=1.12$ than if $\nu=1$. To match the power
  on this scale then requires a higher value of $m_{\rm{WDM}}$ if
  $\nu=1$ than if $\nu=1.12$. We can therefore derive an `equivalent
  $\nu=1.12$' mass for each of our models which gives the best
  approximation to the transfer function in our $\nu=1$ simulations.
  These masses are listed in the final column in Table~\ref{Tab1}. (We
  carry out the comparison for $T^{2}(k)>0.5$ and use the equation
  relating $m_{\rmn{WDM}}$ and $\alpha$ given in Eqn.~7 of 
  \citealt{Viel05}).

  The linear theory power spectra used to set up the initial
  conditions are plotted in Fig.~\ref{PowSpec}. By construction, the
  peak of the power spectrum moves to higher $k$ as $\alpha$ decreases
  (and the particle mass increases). For all WDM models the initial
  power spectrum peaks at a value of $k$ smaller than the Nyquist
  frequency of the particle load in the simulation. This will lead to
  the formation of spurious haloes as mentioned in Section~\ref{intro}.
  
  Self-bound haloes were identified using the \textsc{subfind}
  algorithm \citep{Springel01}; they are required to contain at least
  20 particles. The largest \textsc{subfind} group is the galactic
  halo itself, to which we will refer as the `main halo'. Smaller
  haloes that reside within the main halo are known as `subhaloes',
  whereas those that are outside the main halo are `independent
  haloes'. Most of the subhaloes will have experienced gravitational
  stripping whilst most of the independent haloes will have not.

  \begin{figure}
    \includegraphics[scale=0.33, angle = -90]{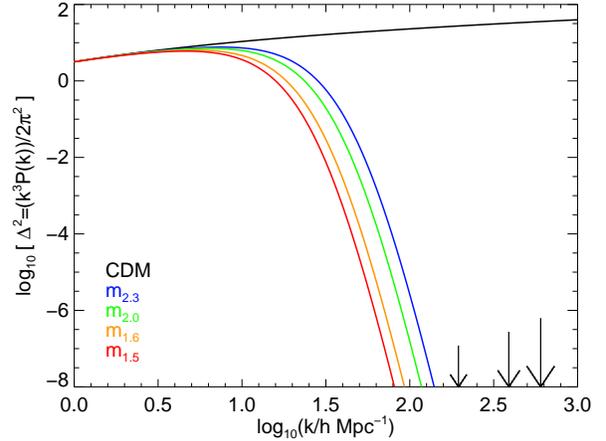}
    \caption{The linear theory power spectrum used in the simulations.
      The black line corresponds to the CDM model, CDM-W7, while the blue,
      green, orange and red lines correspond to the $m_{2.3}$,
      $m_{2.0}$, $m_{1.6}$, and $m_{1.5}$ WDM models respectively.
      The arrows mark, in order of smallest to largest, the Nyquist
      frequency of our low, medium, and high resolution simulations.}
    \label{PowSpec}
  \end{figure}

A first view of the simulations is presented in
Fig.~\ref{Reds0Images}. The smooth component of the main haloes is very
similar in all five models: in all cases, the haloes are similarly
centrally concentrated and elongated. The main difference is in the
abundance of subhaloes. The myriad small subhaloes evident in CDM-W7 are
mostly absent in the WDM models. For these, the number of subhaloes decreases
as $\alpha$ increases (and the WDM particle mass decreases).

  \begin{table}
    \centering
    \begin{tabular}{|c|c|c|c|c|}
      \hline
      Simulation & $M_{200} [\Msun]$ & $r_{200} \rmn{[kpc]}$ & $M_{200b}
      [\Msun]$ & $r_{200b} \rmn{[kpc]}$\\
      \hline  
      CDM-W7 & 1.94$\times10^{12}$ & 256.1 & 2.53$\times10^{12}$ & 432.1\\
      $m_{2.3}$ & 1.87$\times10^{12}$ & 253.4 & 2.52$\times10^{12}$ & 431.4\\
      $m_{2.0}$ & 1.84$\times10^{12}$ & 251.7 & 2.51$\times10^{12}$ & 430.8 \\
      $m_{1.6}$ & 1.80$\times10^{12}$ & 250.1 & 2.49$\times10^{12}$ & 429.9 \\
      $m_{1.5}$ & 1.80$\times10^{12}$ & 249.8 & 2.48$\times10^{12}$ & 429.0 \\
      Aq-A2 & 1.84$\times10^{12}$ & 245.9 & 2.52$\times10^{12}$ & 433.5\\
      \hline
    \end{tabular}
    \caption{Properties of the main friends-of-friends halo in
      each high resolution simulation. The radii $r_{200}$ and $r_{200b}$ 
      enclose regions within which the mean 
      density is 200 times the critical and
      background density respectively. The masses $M_{200}$ and
      $M_{200b}$ are those contained within these radii. We also reproduce data
      from the original Aquarius Aq-A2 halo.}
    \label{Tab2}
  \end{table}

  \begin{figure*}
    \includegraphics[scale=0.35]{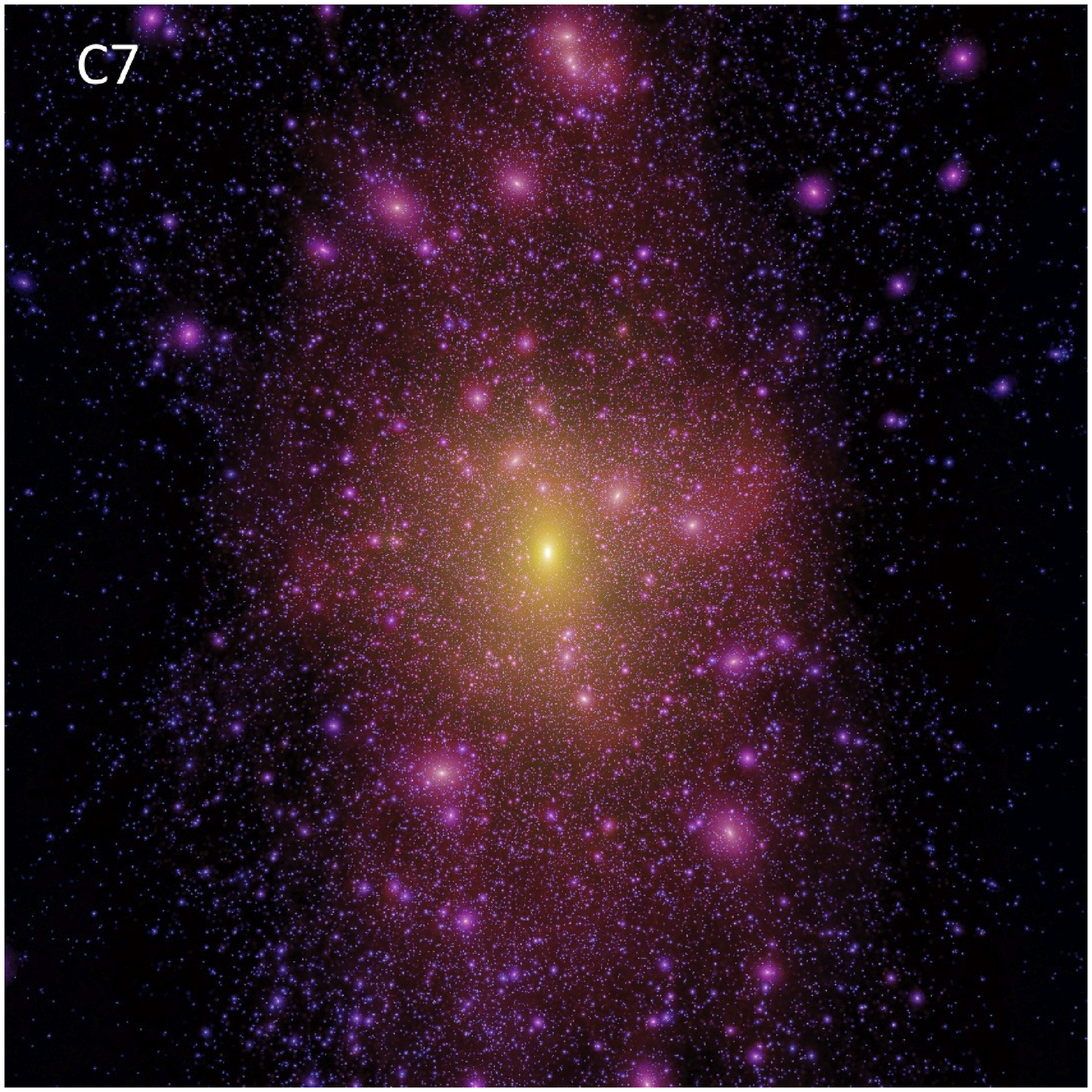}\\
    \includegraphics[scale=0.35]{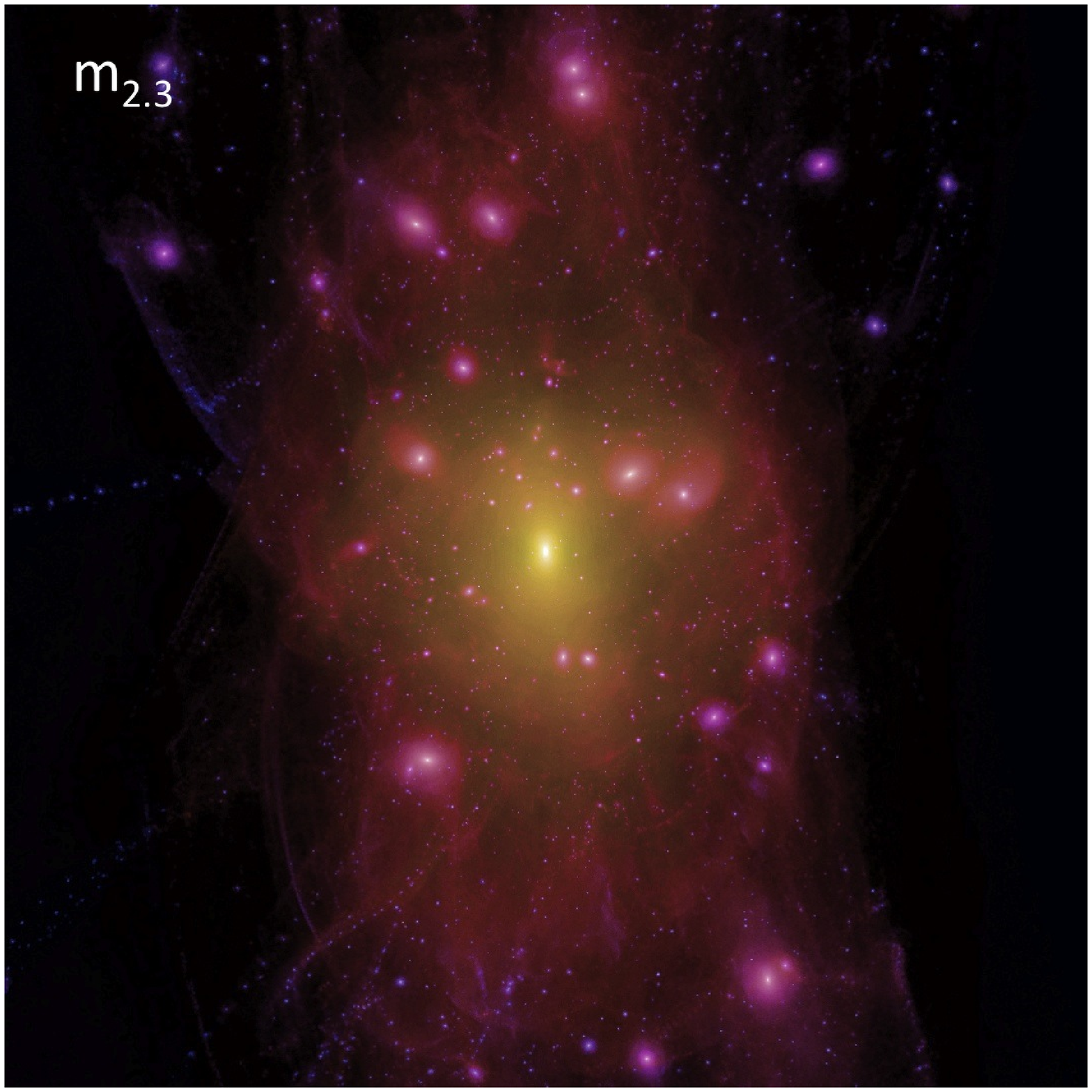}
    \includegraphics[scale=0.35]{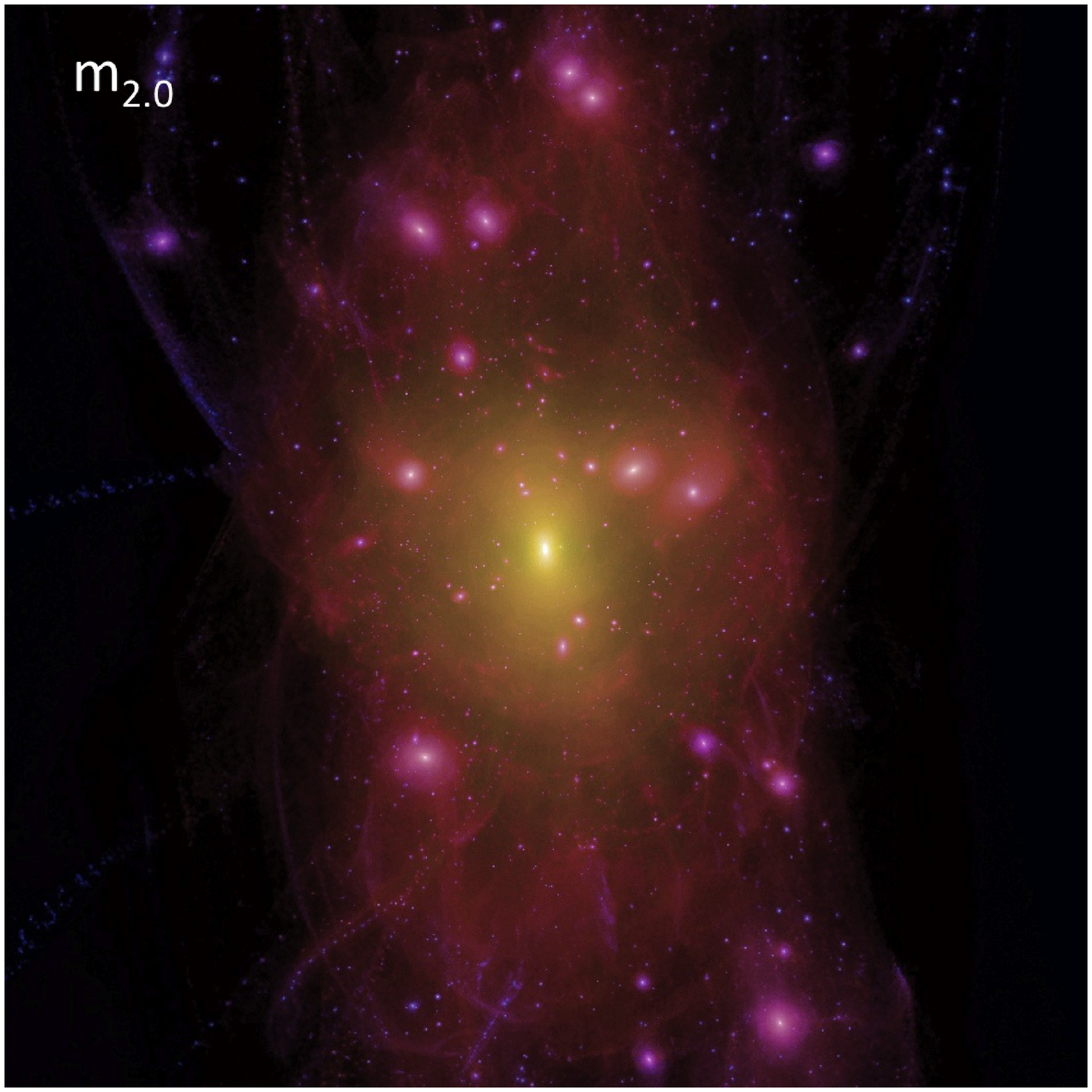}\\
    \includegraphics[scale=0.35]{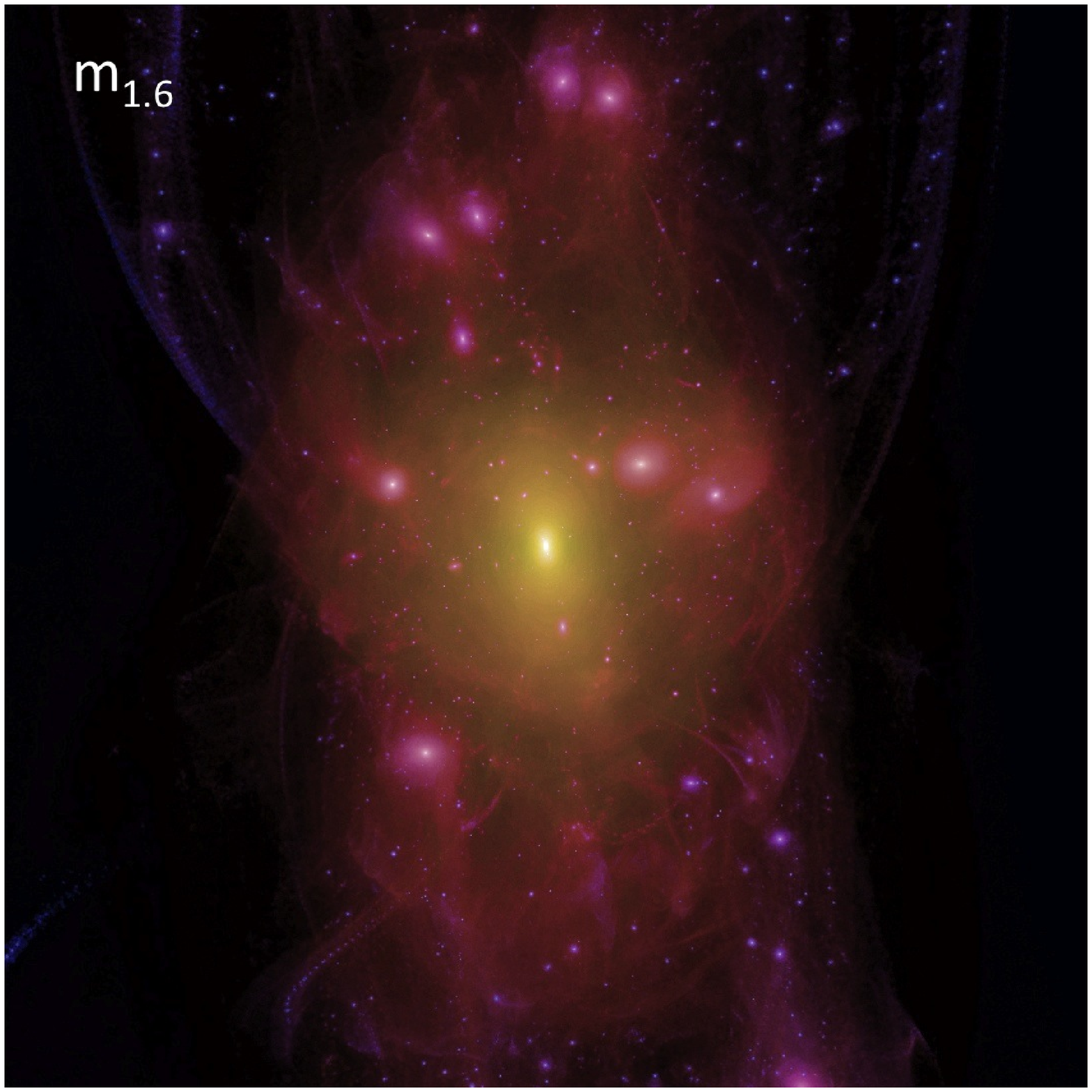}
    \includegraphics[scale=0.35]{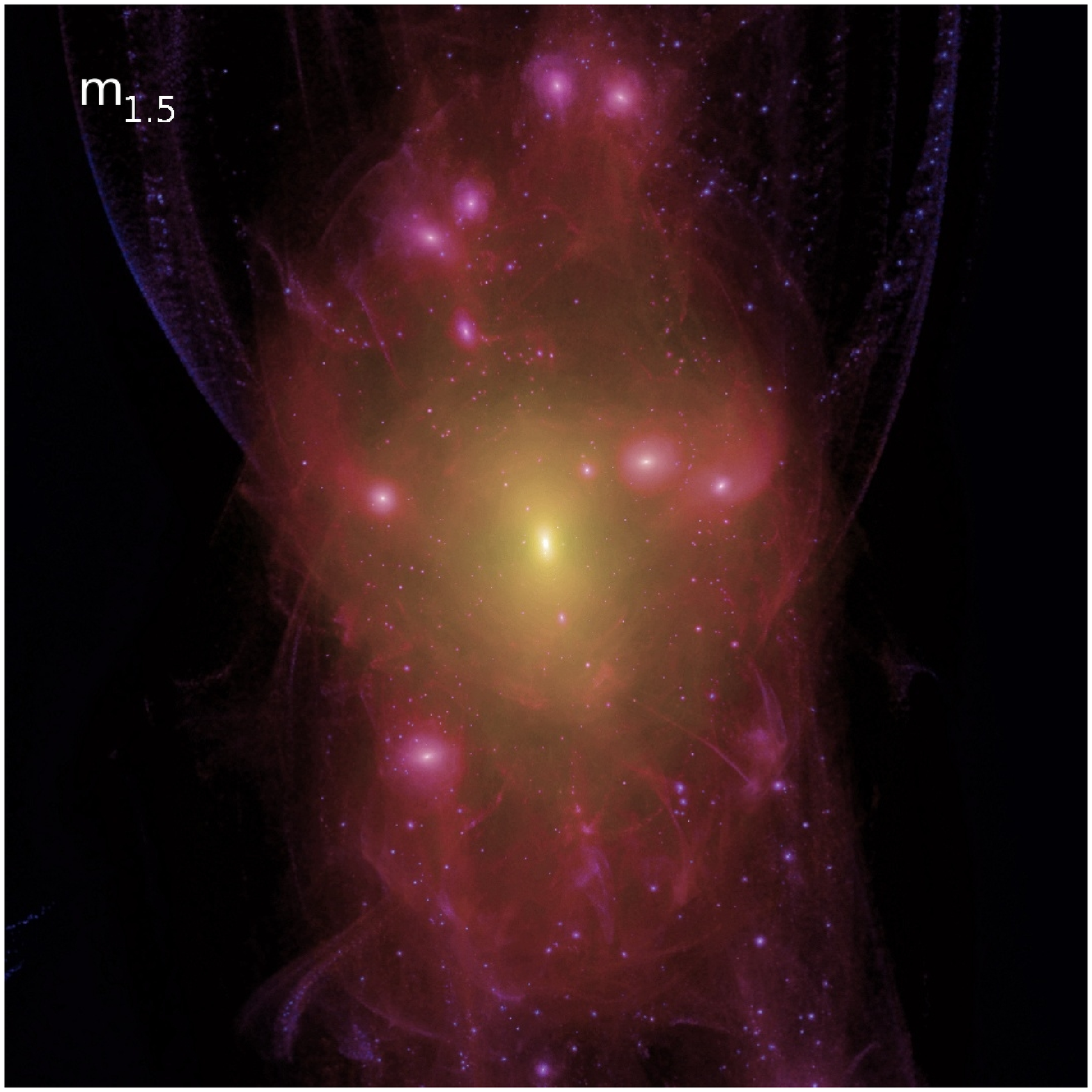}
    \caption{Images of our haloes at redshift $z=0$. The panels show CDM-W7
      (top), $m_{2.3}$, $m_{2.0}$, $m_{1.6}$, and $m_{1.5}$ (left
      to right, then top to bottom). The image intensity and
      hue indicate the projected squared dark matter density
      and the density-weighted mean velocity dispersion
      respectively \citep{Springel08b}. Each panel is 1.5Mpc on a side.}

    
    \label{Reds0Images}
  \end{figure*}
 
  The apparent similarity of the main haloes displayed in
  Fig.~\ref{Reds0Images} is quantified in Table~\ref{Tab2} which lists
  the masses and radii of the largest friends-of-friends halo in each
  simulation. The table gives their masses enclosed within radii of mean
  density 200 times the critical density ($M_{200}$) and 200 times the
  background density ($M_{200b}$). There is a slight trend of
  decreasing mass with increasing $\alpha$, but the maximum change is
  only 7 percent for $M_{200}$ and 2 percent for $M_{200b}$. The change in
  cosmological parameters also makes only a small difference:
  $M_{200}$ is 5 percent higher for CDM-W7 than for the original Aquarius
  halo with WMAP year~1 parameters.

\subsection{The structure of the main haloes}

  The density profiles of the main haloes (including substructures) in
  our high resolution simulations are plotted in
  Fig.~\ref{DensityProf}. There is good agreement amongst all the
  haloes at radii (10-100)~kpc, with the five profiles agreeing to
  better than 10 percent. At larger radii, systematic differences
  between CDM-W7 and the WDM models begin to appear and these become
  increasingly pronounced for the warmer models. These differences are
  due to slight variations in the position of large substructures in
  the outer parts.  There are also small differences at much smaller
  radii ($<10\rmn{kpc}$) which are are not systematic and are thus
  likely due to stochastic variations in the inner regions.

  The radial variation of the logarithmic slope of the density profile
  of each halo is plotted in Fig.~\ref{LogSlope}. In all cases the
  slope at the innermost point plotted approaches the Navarro-Frenk-White (NFW) asymptotic
  value of $-1$ but there is no evidence that the slope is converging.
  There is a slight tendency in the inner parts, $r<4\rmn{kpc}$, for
  the slope in the WDM models to be shallower than in the CDM model,
  but there is no obvious trend with $\alpha$, possibly because of
  stochastic effects in the inner regions.  Thus, apart from minor
  differences, the structure of these $\sim 10^{12}\Msun$ haloes varies
  little with power spectrum cut off, as expected for systems of mass
  $\gg M_{\rmn{th}}$.

  \begin{figure}
    \includegraphics[angle=-90,scale=0.37]{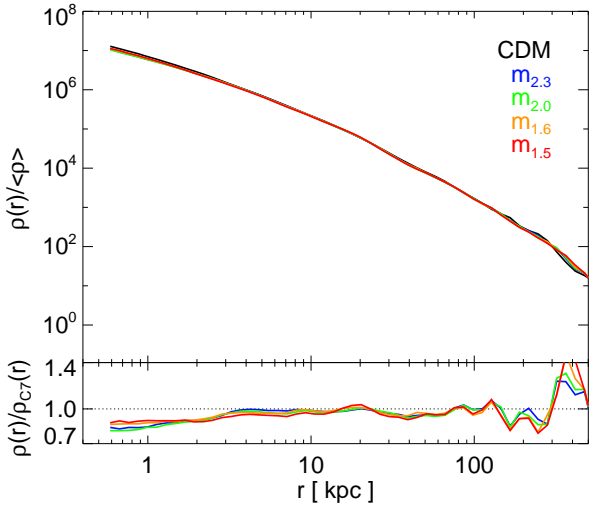}
    \caption{Density profiles of the main haloes (including subhaloes)
      in the simulations normalised by the background matter
      density. The line colours are as in Fig.~\ref{PowSpec}.
      The profiles are plotted only beyond the `Power radius'
      \citep{Power03} at which numerical convergence is expected. The
      bottom panel shows the profiles for the WDM simulations
      normalized to the profile for the CDM-W7 model.}
    \label{DensityProf}
  \end{figure}

  \begin{figure}
    \includegraphics[angle=-90,scale=0.35]{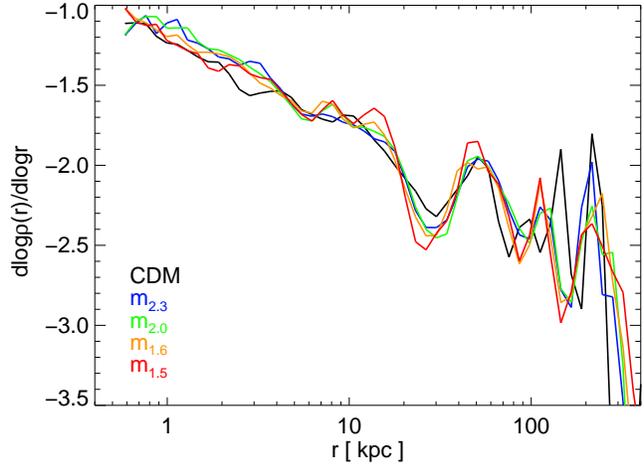}
    \caption{Radial variation of the logarithmic slope of the density
      profiles of the main haloes in the simulations.  Line colours and
      plotting range are as in Fig.~\ref{DensityProf}.}
    \label{LogSlope}
  \end{figure}

  \section{Removal of Spurious Haloes}
  \label{RSSsec}

  One of the main aims of this study is to determine the mass function
  of subhaloes in WDM simulations. However, as we discussed in
  Section~\ref{intro}, simulations in which the initial power spectrum
  has a resolved frequency cutoff can undergo spurious fragmentation
  of filaments. An example is shown in Fig.~\ref{Pos065}, where we
  compare a region in one of our simulations with the corresponding
  region of a higher resolution simulation with the same initial
  conditions by plotting those particles that have collapsed into dark
  matter haloes. In both simulations there are two large haloes and
  several smaller ones. The large haloes have very similar sizes and
  positions in the two simulations, and can be regarded as genuine
  objects. By contrast, the small haloes have different sizes and
  positions in the two simulations; there are also more of them in the
  higher resolution case.  As shown by \cite{Wang07}, increasing the
  resolution even by rather large factors is not sufficient to prevent
  the formation of these artificial haloes. Using glass initial
    conditions, as we do for our simulations, does not reduce this
    problem. Future N-body codes that use phase space smoothing
  techniques may be able to alleviate this problem
  \citep{Hahn13,Shandarin12, Angulo13}. At present, however, the only
  practical measure is to develop a reliable algorithm for identifying
  and removing these `spurious' haloes from the halo catalogues.

  \begin{figure}
    \includegraphics[scale=0.33, angle=-90]{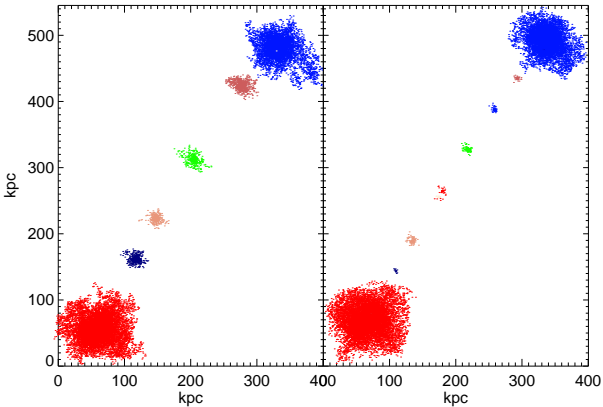}

    \caption{A region of a WDM simulation performed at two different
      resolutions. The particle mass for the high
      resolution simulation (\emph{right}) is 29 times smaller than that of
      the low resolution case (\emph{left}). Only particles in bound
      structures at this snapshot are shown. Particles are coloured
      according to the halo to which they belong. The number
      of particles plotted in each panel is equal to the number of
      bound-structure particles in the low resolution simulation; we
      have applied random sampling in the high resolution case.}
    \label{Pos065}
  \end{figure}

  We now introduce an algorithm for distinguishing between genuine
  and spurious subhaloes. It exploits three properties of the artefacts
  -- mass, resolution dependence and the shape of the initial particle
  distribution -- to define a series of cuts that isolate the
  artefacts. We present an outline of the method in Section~\ref{OotM}
  and provide details in Section~\ref{AtM}.  Note that while the
  results presented here have been derived for subhaloes that have been
  accreted into another halo, the algorithm is equally valid for
  independent haloes.

  \subsection{Outline of the methods}
  \label{OotM}

  Previous simulations have shown that spurious haloes have small
  masses at formation and outnumber genuine haloes on those mass
  scales where they are present \citep{Wang07}. Thus, in principle,
  many spurious haloes can be singled out by applying a mass
  cut. This mass threshold, however, is not well
  defined because the mass function of genuine haloes overlaps that
  of the spurious haloes, so it is useful to introduce additional
  criteria to ensure that, as far as possible, all artificial haloes
  are identified and no genuine ones are removed.

  The resolution dependence of the spurious fragmentation can be used
  to refine the distinction between genuine and artificial haloes.
  While genuine haloes in a simulation at a given resolution are
  expected to be present in the same simulation at higher resolution,
  this need not be the case for spurious haloes, as illustrated in
  Fig.~\ref{Pos065}. \citet{Springel08b} showed that it is possible to
  match haloes and subhaloes between different resolution simulations by
  tracing their particles back to the initial conditions and
  identifying overlapping Lagrangian patches in the two simulations.
  We refer to the initial Lagrangian region of each halo, or more
  precisely the unperturbed simulation particle load, as its
  `protohalo'. The initial positions of the particles displayed in
  Fig.~\ref{Pos065} are shown in Fig.~\ref{PosIC}. The two large
  objects originate from protohaloes of similar size and location, but
  there are clear discrepancies in the number, location and mass of
  the small objects. Thus, attempts to match small haloes in the two
  simulations will often fail because spurious haloes in the low
  resolution calculation do not have a counterpart in the high
  resolution simulation.

  A third criterion exploits the most striking feature visible in
  Fig.~\ref{PosIC}: the shapes of the protohaloes. Genuine protohaloes
  are spheroidal, whereas spurious protohaloes have much
  thinner, disc-like geometries.  They can therefore be easily flagged
  as the progenitors of spurious haloes in the initial conditions.

  \begin{figure}
    \includegraphics[scale=0.45]{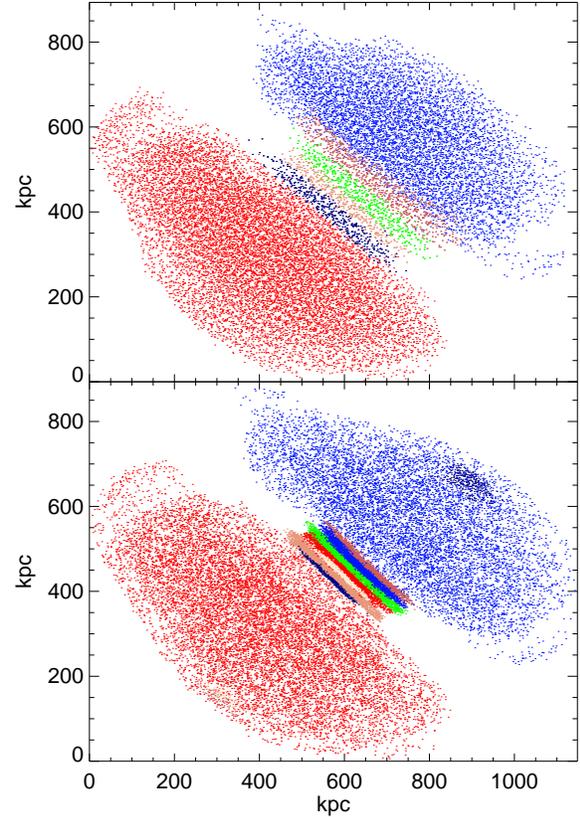}
    \caption{The particles of Fig.~\ref{Pos065} traced back to their
      positions in the initial conditions. The low resolution
      simulation is shown in the top panel and the high resolution
      simulation in the bottom panel. Note the highly flattened
      configurations of spurious haloes.}
    \label{PosIC}
  \end{figure}

  In this study we are interested in objects that become subhaloes at
  the present day. We will apply these three criteria to them in the
  following order. First, we identify a cut based on protohalo
  shape, rejecting from the catalogue all subhaloes flatter than a
  given threshold. Secondly, we apply a mass cut; finally, we refine
  the mass cut using a matching procedure between simulations at
  different resolution. In what follows, we restrict attention to
  subhaloes lying within $r_{200b}$ of the main halo centre at $z=0$
  except where we state otherwise.

  \subsection{Application}
  \label{AtM}
  \subsubsection{Protohalo shapes}
  To determine the flattening of protohaloes we consider all the
  particles that make up a subhalo at some epoch (determined below),
  find their positions in the unperturbed simulation particle
    load and calculate the inertia tensor of the particle set:

  \begin{equation}
    I_{ij}=\sum_{\rmn{all~particles}} m(\delta_{ij}|{\bf x}|^{2}-x_{i}x_{j}),
  \end{equation}

  \noindent
  where $\delta_{ij}$ is the Kronecker delta function, $m$ is the
  particle mass and ${\bf x}$ is the particle position relative to the
  protohalo centre of mass. We take $a\ge b\ge c$ to be the axis
    lengths of the uniform, triaxial ellipsoid that has the same
    moment of inertia tensor as the protohalo. We can then calculate
    $s=c/a$, known as the sphericity. A disc-like (or, more rarely,
  needle-like) spurious subhalo will have a major axis (disc diameter,
  $a$) much longer than its minor axis (disc thickness, $c$), and thus
  a small value of $s$. Genuine subhaloes, on the other hand, are
  spheroidal and thus have higher values of $s$.

  We now need to choose an appropriate epoch at which to identify the
  particles that make up the protohalo. This should be well before the
  subhalo has fallen into a larger halo, after which its outer particles will
  be stripped. We select the earliest simulation
  snapshot below which the halo mass is more than
  half the maximum mass, the `half-maximum mass snapshot'. The initial
  positions of the particles in the object at this time are used to
  evaluate the protohalo sphericity.

  \begin{figure}
    \includegraphics[angle=-90,scale=0.33]{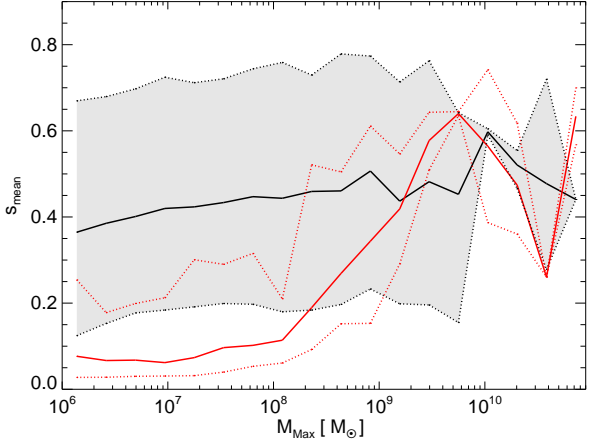}
    \caption{Mean subhalo sphericities as a function of
      $M_{\rmn{Max}}$ for the high resolution CDM-W7 (black) and the
      $m_{1.5}$ (red) runs. The region between the upper and lower 99
      percentiles of the CDM distribution is shown in grey; the same
      region for the $m_{1.5}$ simulation is delineated by the red
      dotted lines.}
    \label{MeanSpher}
  \end{figure}

  The distributions of $s$ for the subhaloes that survive to $z=0$ in
  the CDM-W7 and $m_{1.5}$ simulations are illustrated in
  Fig.~\ref{MeanSpher}, as a function of $M_{\rmn{Max}}$.  The mean
  sphericity is shown as a solid line and the 98 percent range is indicated
  by the dotted lines in each case.  The figure reveals two regimes.
  For values of $M_{\mathrm{Max}}>10^{9}\Msun$, the sphericity
  distributions in the two simulations are consistent with each other.
  For lower masses the protohaloes in the $m_{1.5}$ simulation are
  much flatter than in CDM-W7. This clear dichotomy suggests that most
  of the $m_{1.5}$ subhaloes with $M_{\mathrm{Max}}>10^{9}\Msun$ are
  genuine and most of those with $M_{\mathrm{Max}}<10^{8}\Msun$ are
  spurious. We can use the CDM subhaloes to define a cut in protohalo
  sphericity above which WDM subhaloes are likely to be real.  We find
  that 99 percent of CDM subhaloes containing more than 100 particles
  at the half-maximum mass snapshot have protohaloes with sphericity
  greater than $\sim0.16$ (depending slightly on simulation
  resolution), which we denote \scut. We exclude from our cleaned
  subhalo catalogue any WDM subhalo whose protohalo has sphericity
  less than \scut, regardless of mass.  This cut rejects between 86
  percent ($m_{2.3}$) and 93 percent ($m_{1.5}$) of the WDM
  subhaloes as spurious. We have checked, as we show later, that the
  subhaloes rejected by this criterion do not have clear counterparts
  in pairs of simulations of different resolution, where in this case
  the difference in resolution is a factor of 8. We find that
    varying \scut by 20 percent changes the number of subhaloes
    identified as genuine by less than 20 percent, which is within
    the $2\sigma$ Poisson uncertainty in the number identified using
    our chosen value of $s_{\rm cut}$.

  \subsubsection{A first guess of the mass cut}
  For a first guess of the mass cut below which a majority of
  subhaloes are spurious, we resort to the results of \citet{Wang07}. They
  showed that the characteristic mass below which spurious subhaloes begin
  to dominate the subhalo mass function is related to the matter power
  spectrum cutoff and the simulation resolution. The larger the value
  of the cutoff frequency and the higher the resolution of the
  simulation, the smaller is the mass of the largest spurious
  subhaloes. \citet{Wang07} derived an empirical formula for the mass
  at which spurious subhaloes begin to dominate:

  \begin{equation}
    M_{\mathrm{lim}} = 10.1\bar\rho dk_{\mathrm{peak}}^{-2},
  \end{equation}

  \noindent
  where $\bar\rho$ is the mean density of the Universe, $d$ is the
  mean interparticle separation (a measure of resolution), and
  $k_{\mathrm{peak}}$ is the wavenumber at which the dimensionless
  power spectrum, $\Delta^2(k)$, has its greatest amplitude.  We can
  apply this formula to $M_{\rmn{Max}}$ to estimate a cut below which
  the majority of the subhaloes will be spurious. Some genuine haloes
  will have $M_{\rmn{Max}}$ below this threshold but the mass limit
  can be refined using the matching criterion.

  \subsubsection{Matching subhaloes between simulations}
  A subhalo that is present in both a low resolution simulation (LRS)
  and in its high resolution counterpart (HRS) is likely to be
  genuine. We can use this property to refine the mass cut. We set the
  cutoff mass to be $M_{\mathrm{min}} = \kappa M_{\mathrm{lim}}$, where
  $\kappa$ is a constant such that the number of LRS subhaloes of mass
  greater than $M_{\rmn{min}}$ is equal to the number of subhaloes with
  matches in the HRS. We will assume that the value of $\kappa$
  determined for the LRS subhaloes is also applicable to the HRS
  catalogues.

  We now introduce an algorithm for finding high resolution
  counterparts of the low resolution subhaloes. Genuine haloes should
  originate from the same Lagrangian region regardless of resolution.
  Therefore, to match subhaloes we require a quantitative measure
  to compare these Lagrangian regions in simulations of different
  resolution and check that they overlap and have the same
  shape. These shapes are defined by point-like particles. In order to
  develop a quantitative measure of the overlap we need to smooth these points. We measure
  the degree to which a pair of objects in different
  resolution simulations are the `same' by comparing the entirety of
  the regions from which they form. We introduce a statistic:

  \begin{equation}
    R = \frac{U_{\rmn{AB}}^2}{U_{\rmn{AA}}U_{\rmn{BB}}},
  \end{equation}

  \noindent
  where $U_{XY}=\int\phi_{\rmn{X}}\rho_{\rmn{Y}}dV$, $V$ is volume,
  and $\rho_{\rmn{A/B}}$ and $\phi_{\rmn{A/B}}$ are the density of and
  gravitational potential due to the matter distributions A/B
  respectively. It can be shown using Green's Theorem that if the
  matter distribution of subhalo A is proportional everywhere to that
  of subhalo B, $R=1$; for any other configuration $R<1$. We apply
  this formula to our candidate LRS-HRS protohalo particle
  distributions, representing each particle as a spherical shell of
  radius equal to the LRS mean interparticle separation and with
  infinitesimal thickness. The best match for the LRS subhalo will
  then be the HRS halo with which it attained the highest value of
  $R$. We retain this value of $R$ for each LRS subhalo as our measure
  of its matching quality. A genuine LRS subhalo will have a good
  match at high resolution and therefore have a value of $R$ close to
  1, whereas a spurious subhalo will have a poor match and a lower
  value of $R$.

  To find candidate matches, we first divide the simulation volume
  into a grid of cells of comoving length $\gsim60\rmn{kpc}$, and, for
  a given low resolution protohalo, choose as candidate matches the
  high resolution protohaloes that occupy the same and neighbouring
  grid cells. It is computationally expensive to calculate $R$ for the
  largest subhaloes, but we found that random sampling of each halo
  with 10000 particles returned values of $R$ that did not vary
  systematically with $M_{\rmn{Max}}$ for subhaloes of
  $M_{\rmn{Max}}>10^{9}\Msun$. We therefore adopt a threshold of 10000
  particles. When attempting to match subhaloes between simulations,
  minor differences in which particles are assigned to each subhalo
  can have an impact on $R$. We mitigate this problem by performing
  the calculation for both the maximum-mass and half-maximum mass
  snapshots, selecting the higher value of the two for each
  subhalo. The resulting values of $R$ are plotted as a function of
  $M_{\rmn{Max}}$ in Fig.~\ref{MDP}.

  \begin{figure}
    \includegraphics[angle=-90,scale=0.33]{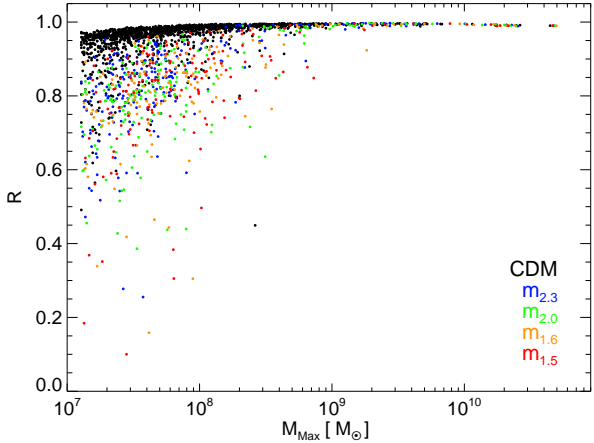}
    \caption{$R$ as a function of $M_{\rmn{Max}}$ for CDM and WDM LRS
      subhaloes matched to HRS counterparts (those that fail the
      sphericity cut are still included). The black dots denote CDM
      subhaloes, blue $m_{2.3}$, green $m_{2.0}$, orange $m_{1.6}$, and
      red $m_{1.5}$ (the same as Fig.~\ref{PowSpec}.}  \label{MDP}
  \end{figure} 

At high masses, the CDM and WDM protohaloes have $R$
  close to 1.  As the protohalo mass decreases, $R$ becomes
  systematically lower and the decline is much steeper for the WDM
  models, as expected in the presence of poorly matching spurious
  subhaloes.  Unfortunately, a small proportion of CDM subhaloes
  also attain low values of $R$ and the demarcation between the
  distributions of $R$ for WDM and CDM is much less clear cut than we
  found for the sphericity measurement, $s$. Were we to take the same
  approach for $R$ as we did for $s$, we would infer a cut in $R$ of
  about 0.68. More than half of the WDM subhaloes have a value of $R$
  closer to 1 than this, and since the sphericity-based algorithm
  rejects $\sim90$ percent of subhaloes, adopting this cut in $R$ would
  return a heavily contaminated sample.  We circumvent this problem by
  using our sphericity cut to determine the distribution of $R$ for
  spurious subhaloes. For each WDM model, we take 10000 subsamples of
  100 subhaloes that fail the sphericity cut (with replacement) and
  take the second highest $R$ of each subsample to be the threshold,
  $R_{\rmn{min}}$, below which subhaloes are spurious. This result
    is not sensitive to the size of our subsamples. The mean value of
  $R_{\rmn{min}}$ across the 10000 subsamples is found to be in the
  range 0.94-0.96 for each of the four WDM models.  For those
    subhaloes that instead pass the sphericity cut, the mean value of
    $R_{\rmn{min}}$ is greater than 0.995 for all four models, showing
    that sphericity is a robust and accurate diagnostic of whether or
    not an object is spurious.  

    We now couple the matching and
  sphericity criteria to determine the optimal cut in $M_{\rmn{Max}}$.
  In Fig.~\ref{MvSL4}, we plot $s$ as a function of $M_{\rmn{Max}}$
  for the LRS subhaloes in each of our four WDM models, indicating
  their matching quality by colour.  We adopt $R_{\rmn{min}}=0.94$. We
  restrict attention to subhaloes that pass the sphericity cut and take
  a mass limit $M_{\rmn{min}}=\kappa M_{\rmn{lim}}$ such that the
  number of subhaloes with mass greater than $M_{\rmn{min}}$ is equal
  to the number of subhaloes with $R>R_{\rmn{min}}$. In
  Fig.~\ref{MvSL4} this is equivalent to the number of red dots to the
  right of the mass cut being equal to the number of blue dots to the
  left.  We find that this condition requires values of $\kappa$
  between 0.4 and 0.6, given the uncertainty in $R_{\rmn{min}}$. For
  simplicity, we will adopt $\kappa=0.5$; we find that this value
  provides a good compromise between rejecting low mass genuine
  objects and including high mass spurious subhaloes in all four
  models.  Varying $R_{\rmn{min}}$ and $\kappa$ in the range stated
  here makes a difference of $\sim10$ percent to the number of
  subhaloes returned in the $m_{1.5}$ model and $\sim5$ percent in the
  other cases. The values of $M_{\rmn{min}}$ are then
  $1.5\times10^8\Msun$, $2.2\times10^8\Msun$, $3.2\times10^8\Msun$,
  and $4.2\times10^8\Msun$ for the $m_{2.3}$, $m_{2.0}$, $m_{1.6}$,
  and $m_{1.5}$ models respectively in the low resolution simulations.
  For the high resolution simulations, they decrease to
  $5.1\times10^{7}\Msun$, $7.0\times10^{7}\Msun$,
  $1.1\times10^{8}\Msun$, and $1.4\times10^{8}\Msun$.  

  \begin{figure}
    \includegraphics[scale=0.40]{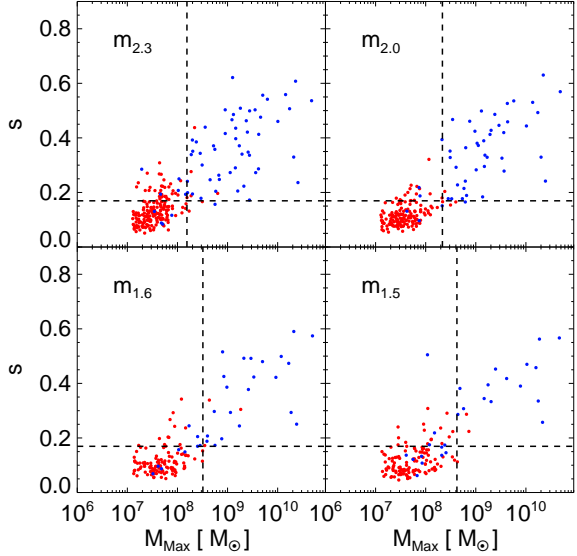} \caption{Dot
      plots of $s$ and $M_{\rmn{Max}}$ for subhaloes in the four
      different WDM models at low resolution. Blue points correspond
      to $R\ge 0.94$ and red points to $R<0.94$. The horizontal,
      dashed line is \scut\ and the vertical line is
      $M_{\mathrm{min}}$. All subhaloes are within $r_{200b}$ of the
      main subhalo centre at redshift zero.}  \label{MvSL4}
  \end{figure} 

  To summarize, we have used the mass, resolution dependent, and
  Lagrangian region shape properties to identify spurious subhaloes in
  our subhalo catalogues. Having derived values for \scut\ and $M_{\rmn{min}}$ --
  the latter as a function of power spectrum cutoff and resolution --
  we can apply these cuts to the high resolution simulations. We plot
  the results in Fig.~\ref{MvSL2}. Changing the value of $\kappa$ in
  the range 0.4-0.6 produces a variation of $<5$ percent in all four
  HRS models, and this does not affect our conclusions. In what
  follows we consider only those subhaloes that pass the cuts in each
  of these panels.  

  \begin{figure}
    \includegraphics[scale=0.40]{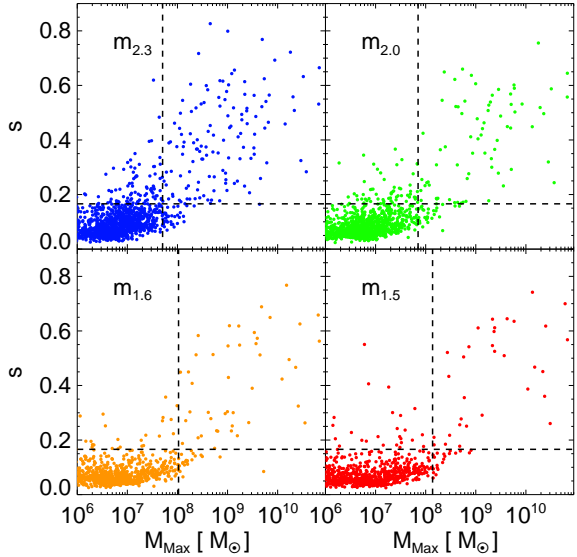} \caption{Dot
      plots of $s$ and $M_{\rmn{Max}}$ for subhaloes in the four
      different WDM models at high resolution. The horizontal, dashed
      line is \scut\ and the vertical line is $M_{\mathrm{min}}$. All
      subhaloes are within $r_{200b}$ of the main subhalo centre at
      redshift zero.}  \label{MvSL2} 
  \end{figure} 

  \section{Results}
  \label{Res} 

  \subsection{The subhalo mass and \vmax functions}
  \label{MassVmax} 

  In Fig.~\ref{CmlVmaxAb} we present the cumulative
  distributions of subhalo mass, $M_{\rmn{sub}}$, and \vmax at $z=0$,
  where \vmax is defined as the peak amplitude of the circular
  velocity profile $V_{\rmn{circ}} = \sqrt{GM(<r)/r}$, with $G$ the
  gravitational constant and $M(<r)$ the mass enclosed within radius
  $r$. This is a useful proxy for mass that is insensitive to the
  definition of the edge of the subhalo. The figure includes both
  genuine (solid lines) and spurious (dashed lines) subhaloes. Overall,
  the spurious subhaloes outnumber the genuine ones by a factor of 10.
  However, the mass function is dominated by genuine haloes beyond
  $M_{\rmn{sub}}\sim (1-3)\times 10^7\Msun$, corresponding to
  \vmax$\sim (4-6)$~\kms, for the different models.  The differential
  mass function (relative to the CDM mass function) for genuine haloes
  in the $m_{2.3}$ case can be fit with the functional form given by
  \citet{Schneider2012}: 

  \begin{equation} 
    n_{\rmn{WDM}}/n_{\rmn{CDM}} = (1+M_{\rmn{hm}}M^{-1})^{\beta}, 
  \end{equation} 
  
  \noindent where
  $M_{\rmn{hm}}$ is the mass associated with the scale at which the
  WDM matter power spectrum is suppressed by 50 percent relative to the CDM
  power spectrum, $M$ is subhalo mass and $\beta$ is a free parameter.
  The best fit value is $\beta$ of 1.3, slightly higher than the value
  of 1.16 found by \citet{Schneider2012} for friends-of-friends haloes
  (rather than \textsc{subfind} subhaloes as in our case). A slightly
  better fit is obtained by introducing an additional parameter,
  $\gamma$, such that: 

  \begin{equation} 
    n_{\rmn{WDM}}/n_{\rmn{CDM}} = (1+\gamma M_{\rmn{hm}}M^{-1})^{\beta}, 
  \end{equation} 

  \noindent with $\gamma=2.7$ and $\beta=0.99$.  However, better statistics are
  required to probe the subhalo mass function more precisely.  

  In principle, comparison of the abundance of subhaloes shown in
  Fig.~\ref{CmlVmaxAb} with the population of satellite galaxies
  observed in the Milky Way can set a strong constraint on the mass of
  viable WDM particle candidates. Assuming that every
  satellite possesses its own dark matter halo and that the parent
  halo in our simulations has a mass comparable to that of the Milky
  Way halo, a minimum requirement is that the number of subhaloes in
  the simulations above some value of $M_{\rmn{sub}}$ or \vmax should
  exceed the number of Milky Way satellite above these values. In
  practice, the comparison is not straightforward because: {\em
    (i)}~the values of $M_{\rmn{sub}}$ or \vmax for the observed
  population are not well known and {\em (ii)}~the total number of
  Milky Way satellites is uncertain. Nevertheless, we can obtain a
  conservative limit on the mass of the particle as follows.  There
  are 22 satellites in the Milky Way for which good quality
  kinematical data exist \citep{Walker09,Wolf10}. Eleven of these are
  `classical satellites' and the remainder are SDSS satellites. Of
  the classical satellites, eight are dwarf spheroidals and the others
  are the large and small Magellanic clouds (LMC and SMC) and
  Sagittarius. \cite{Wolf10} have estimated values of the mass (and
  line-of-sight velocity dispersion, $\sigma^2_{\rmn{los}}$) within
  the (deprojected 3D) half-light radius for the eight classical and
  11 SDSS dwarf spheroidals. These are essentially insensitive to the
  velocity anisotropy of the stellar populations. The circular
  velocity within this radius is then given by: 

  \begin{equation}
    V_{\rmn{circ}}(r_{1/2})=\sqrt{3 \sigma^2_{\rmn{los}}}.
    \label{eqn:vcirc} 
  \end{equation} 
  
  \noindent The values of
  $V_{\rmn{circ}}$ are lower limits to \vmax for each satellite.
  Leo~IV has the smallest circular velocity,
    $V_{\rmn{circ}}=5.7\pm2.9$~\kms, of the 22 studied by \cite{Wolf10}.
  We show in Appendix~\ref{ConStudy} that our simulations have
    converged to better than 8 percent at this value of \vmax, showing that
    our conclusions are not affected by resolution issues
    \citep[c.f.][]{Polisensky2011}.  As shown by \cite{Springel08a},
  values of \vmax for subhaloes in Aquarius level~2 simulations are
  converged to within $\sim 10$ percent for \vmax $\geq 1.5$~\kms.
  We have examined the convergence in our $m_{2.3}$ model and
    find that our L3 and L2 resolution \vmax functions are converged
    to within $2\sigma$ (Poisson) of each other for \vmax$>4$~\kms.
    This is more modest than for the CDM Aquarius simulations, but
    sufficient to resolve the Leo~IV type satellites. This result also
    gives us confidence that our ability to count satellites is not
    impaired by the numerical issues \citep[c.f.][]{Polisensky2011}.

  The known number of satellites in the Milky Way halo, 22, is a lower
  limit to the total number within 280~kpc of the galaxy's centre, the
  distance to which the tip of the red giant branch can be detected in
  the SDSS. This is because although all the classical satellites
  (i.e.  satellites brighter than $M_V=-11$) have probably been
  discovered, SDSS surveyed only 20 percent of the sky [data release 5(DR5)].  Thus, a
  conservative lower limit to the WDM particle mass
  is obtained by requiring that the simulation should produce at least
  22 satellites within this radius with \vmax$>5.7$~\kms. Our
  $m_{1.5}$ simulation produced only 25 subhaloes with \vmax
  greater than this value within the larger radius,
  $r_{200b}=429$~kpc. Furthermore, the mass of the $m_{1.5}$ halo,
  $M_{200}=1.80\times 10^{12}\Msun$, is towards the higher end of
  acceptable values for the mass of the Milky halo; simulations of
  haloes with lower mass would produce even fewer subhaloes. Finally,
  any residual contamination by spurious subhaloes would artificially
  inflate the numbers in our subhalo sample. Thus, we can safely set a
  conservative lower limit to the mass of the WDM particle
  of $m_{\rmn{WDM}}=1.5$~keV.  

  We can set a less conservative but
  still robust lower limit to $m_{\rmn{WDM}}$ by correcting the
  observed number of SDSS satellites to take into account the area
  surveyed. A simple extrapolation multiplying the observed number by
  a factor of 5 has to be taken with caution because we know that the
  classical satellites are not distributed isotropically but are
  concentrated towards a plane, called the `Great pancake' by
  \cite{Libeskind05}.  However, from analysis of the Aquarius
  simulations, \cite{Wang12} have argued that such flat configurations
  occur only for the most massive $\sim 10$ subhaloes and the
  anisotropy of the distribution falls off rapidly with increasing
  sample size so that samples of $\sim 50$ subhaloes follow quite close
  the overall shape of the halo.  Based on this, we do not make any
  corrections for anisotropy and conclude that the Milky Way contains
  at least $11+5\times 11= 66$ satellites with \vmax$>5.7$\kms within
  280~kpc. Using the same argument as before, counting out to a radius
  of 419~kpc in the simulations to be conservative, we find that only
  the $m_{2.3}$ and CDM models produces enough satellites to satisfy the
    limit.  

   To make an estimate of the halo-to-halo scatter, we make use of the
   result of \citet{BoylanKolchin10} that the intrinsic scatter in the
   abundance of CDM subhaloes, $\sigma_{\rmn{scatter}}$, can be fit by
   the sum of the Poisson, $\sigma^{2}_{\rmn{P}}$, and intrinsic,
   $\sigma^{2}_{\rmn{I}}$, variances:

\begin{equation}
  \sigma^{2}_{\rmn{scatter}} = \sigma^{2}_{\rmn{P}}+\sigma^{2}_{\rmn{I}},
\end{equation}

\noindent where $\sigma^{2}_{\rmn{P}}=\langle N\rangle$ and
  $\sigma^{2}_{\rmn{I}}=s_{\rmn{I}}\langle N\rangle^{2}$. Here, $s_{\rmn{I}}$ is
  a constant, which \citet{BoylanKolchin10} calibrate against their
  simulation results and thus obtain  $s_{\rmn{I}}=0.18$. They also found that the
  probability distribution for the
  number of subhaloes $N$, given the mean $\langle N\rangle$ and
  intrinsic coefficient
  $s_{\rmn{I}}$, is well described by the negative binomial distribution:

\begin{equation}
  P(N|r,p) = \frac{\Gamma(N+r)}{\Gamma(r)\Gamma(N+1)}p^{r}(1-p)^{N},
\end{equation} 

\noindent where $p=[1+s_{\rmn{I}}^2\langle N\rangle]^{-1}$ and
$r=s_{\rmn{I}}^{-2}$. We then adopt the number of subhaloes within
$r_{200b}$ from each of our models as the distribution mean and
compute the probability that a given halo will have at least 66
subhaloes. This probability equals 22 percent for $m_{2.0}$ and 0.30
percent for $m_{1.6}$. Therefore, we conclude on this evidence that
$m_{\rmn{WDM}}>1.6$~keV\footnote{ To check whether this limit is
  sensitive to our choice of \scut, we repeated the analysis lowering
  \scut by 20 percent. In this case the probability for the $m_{1.6}$
  model increases to 2.7 percent; thus this mass is still excluded at
  95 percent confidence.}}. This is a more conservative limit than
found by \citet{Polisensky2011}, although our choice of central halo
is slightly more massive than theirs. A larger suite of WDM
simulations is required to determine more precisely the variation in
WDM subhalo abundance at a given host halo mass as well as the
systematic variation of abundance with host halo mass.

\begin{figure}
    \includegraphics[angle=-90,scale=0.33]{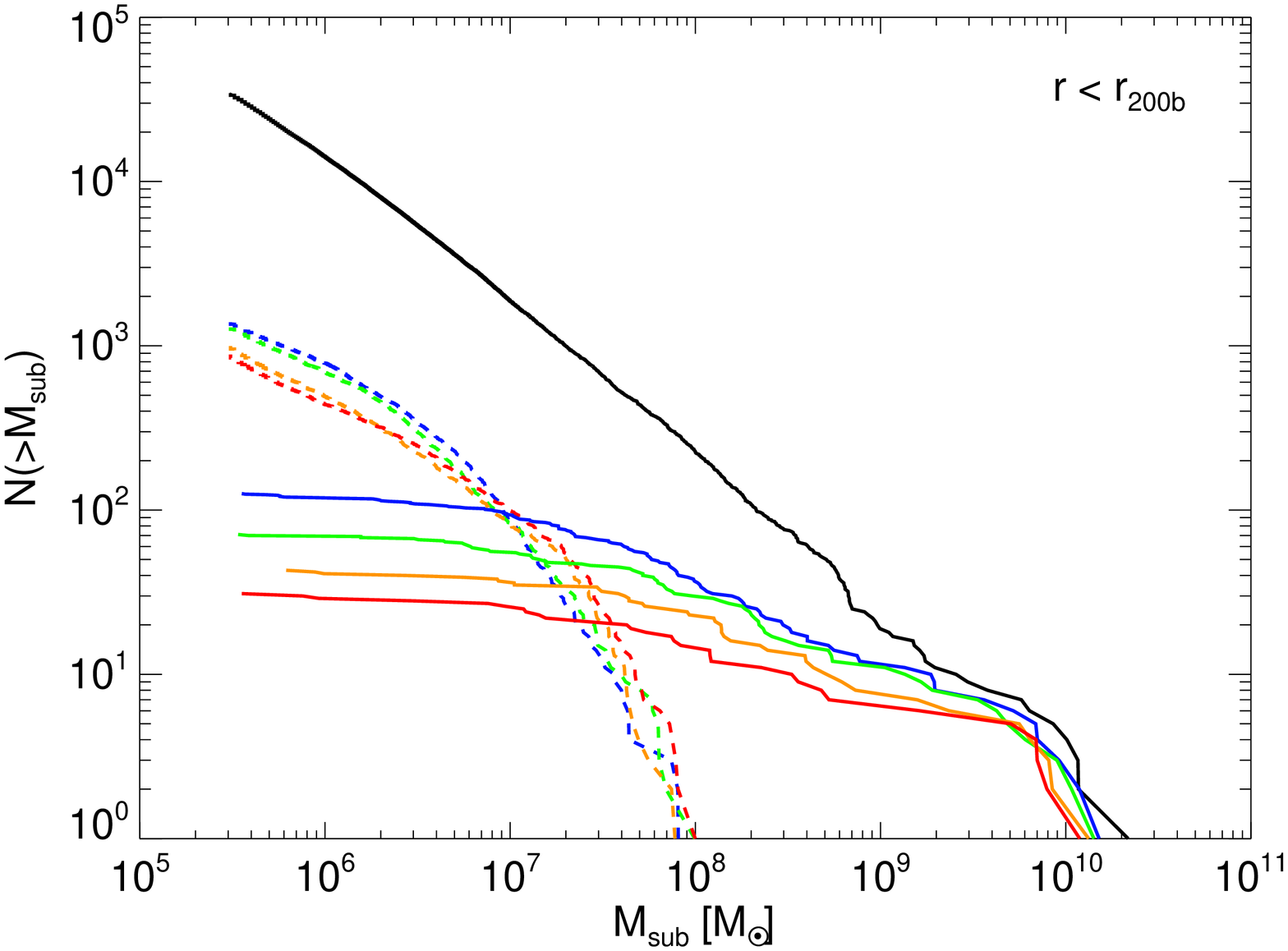}
    \includegraphics[angle=-90,scale=0.33]{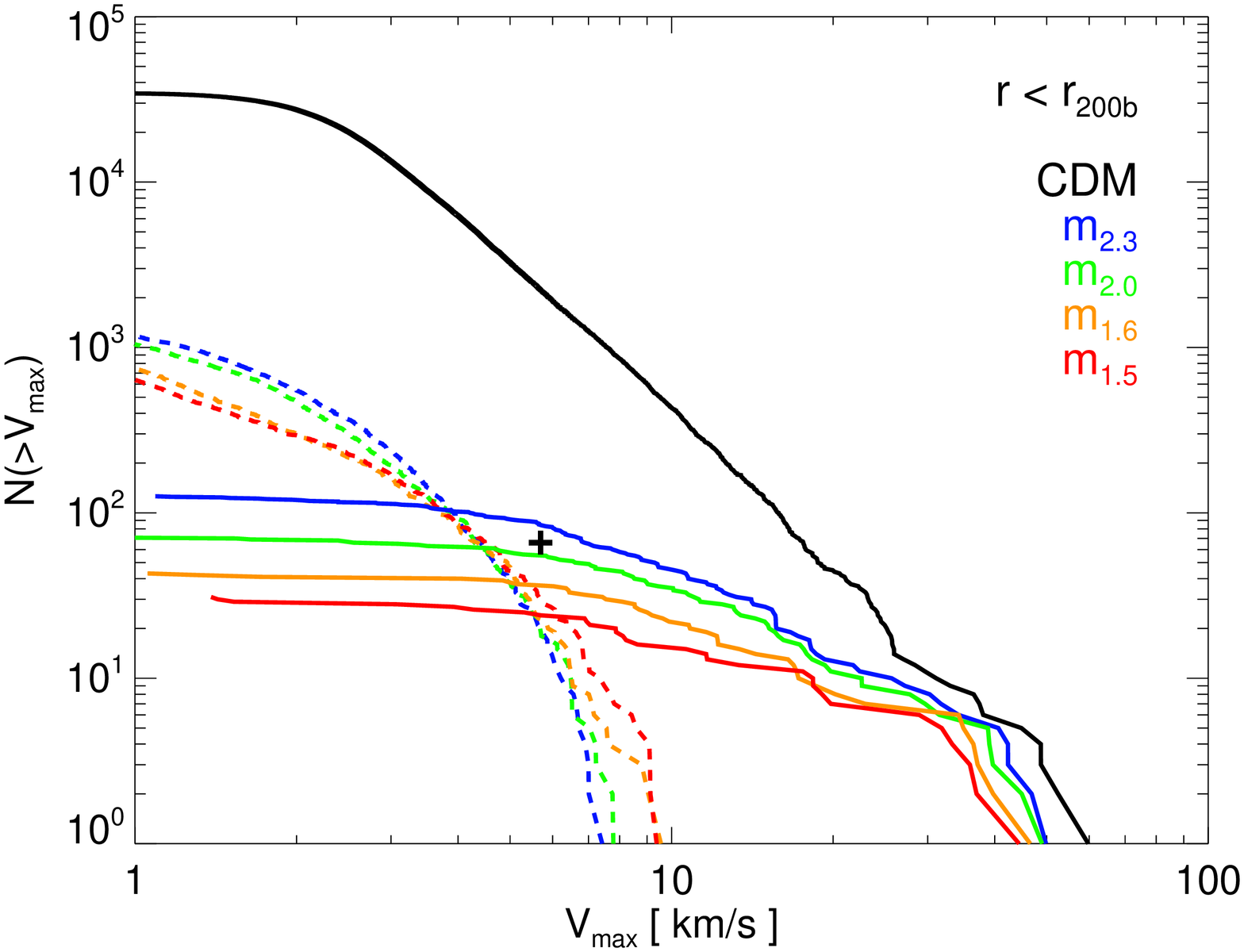}
    \caption{Cumulative subhalo mass, $M_{\rmn{sub}}$, (\emph{top
        panel}) and \vmax (\emph{bottom panel}) functions of subhaloes
      within $r<r_{200b}$ of the main halo centre in the high
      resolution simulations at $z=0$. Solid lines correspond to
      genuine subhaloes and dashed lines to spurious subhaloes. The
      black line shows results for CDM-W7 and the colours lines for
      the WDM models, as in Fig.~\ref{PowSpec}. The black cross in the
        lower panel indicates the expected number of satellites of
        \vmax$>5.7$\kms as derived in the text.}  
     \label{CmlVmaxAb}
  \end{figure}

\subsection{The radial distribution of subhaloes} 

\begin{figure}
    \includegraphics[angle=-90,scale=0.33]{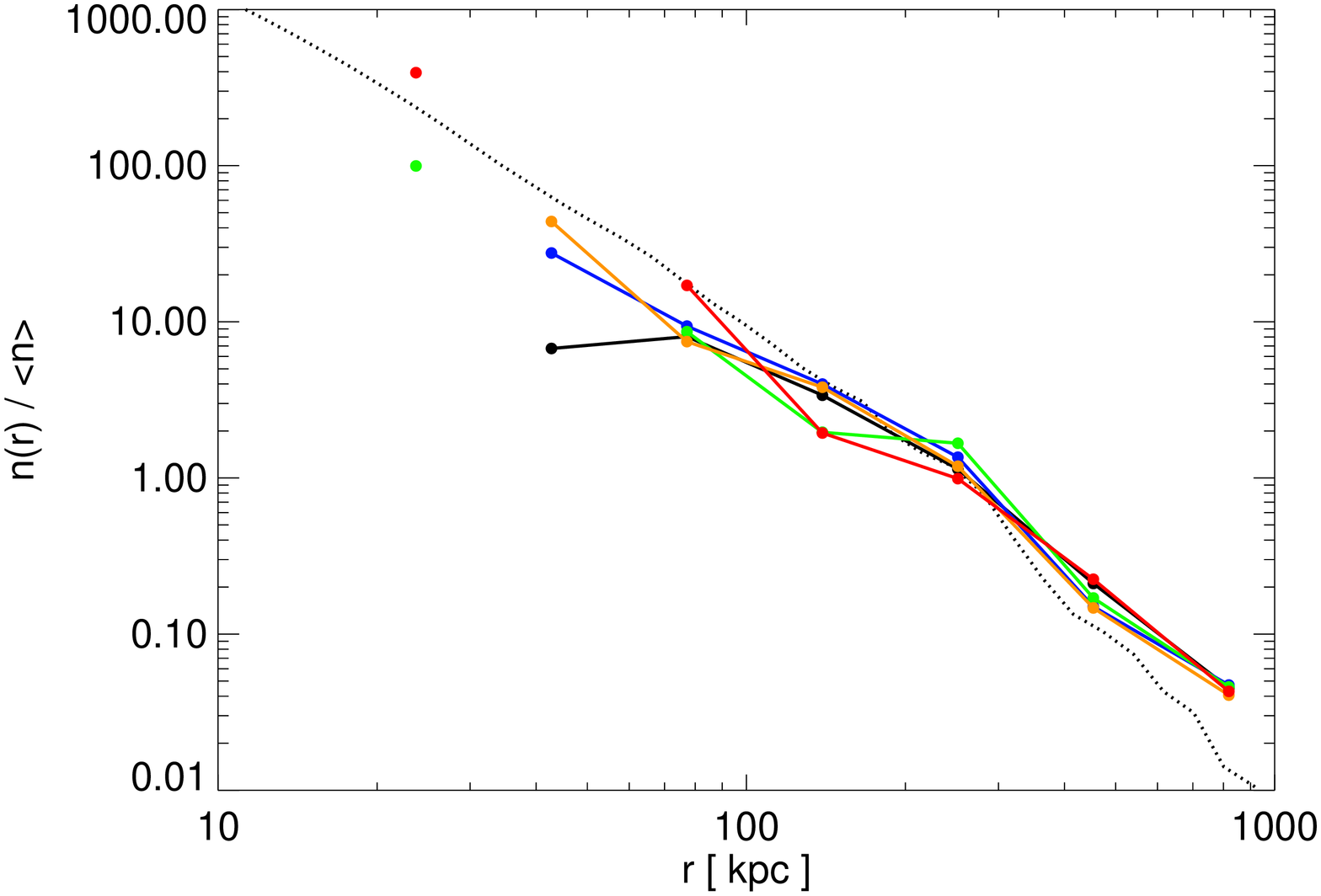}
    \includegraphics[angle=-90,scale=0.33]{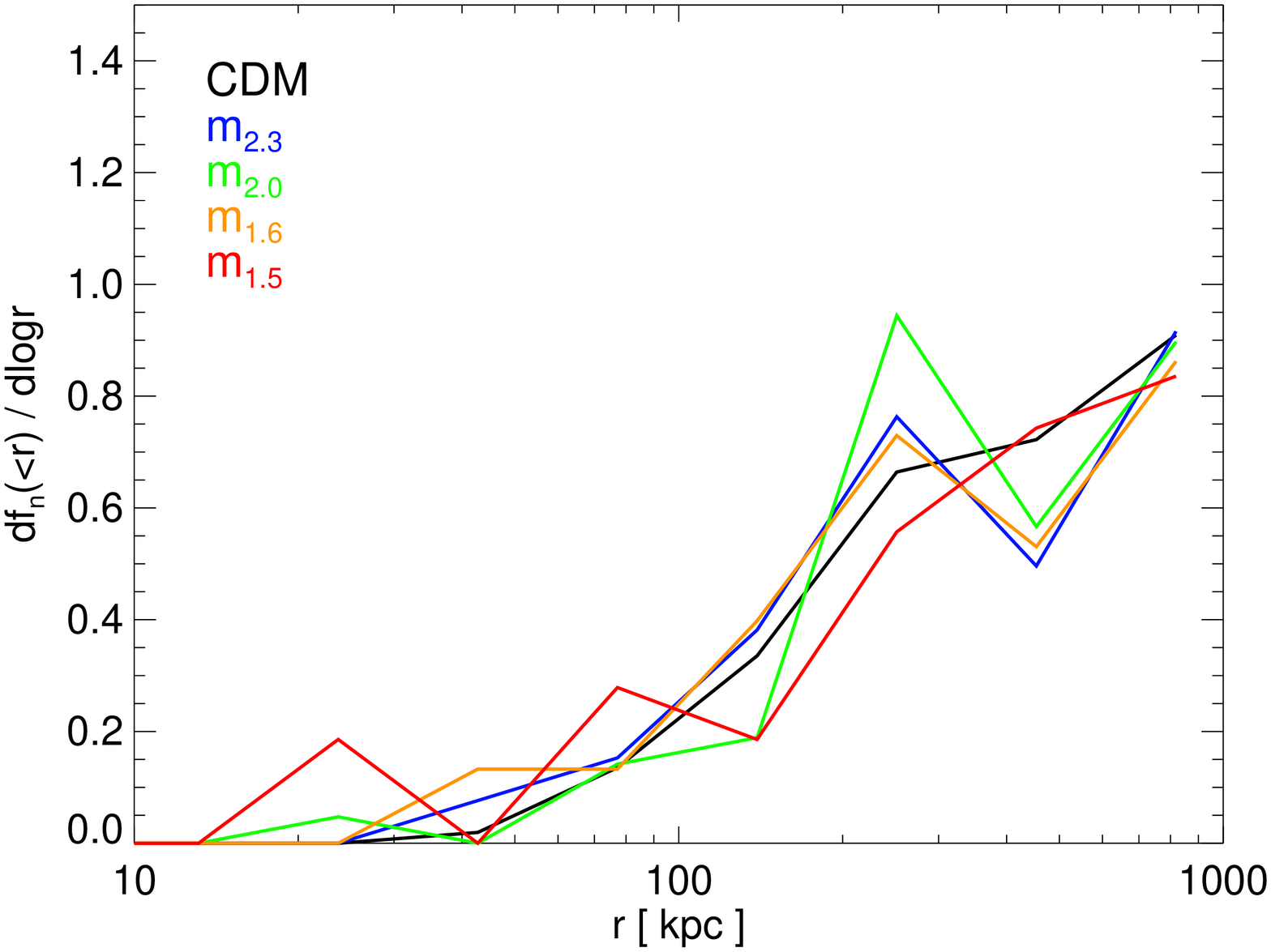}
    \caption{The radial distribution of subhaloes. \emph{Top:} the
      spherically averaged number density of
      $M_{\rmn{sub}}>10^{8}\Msun$ subhaloes normalised to the mean
      overdensity at $r_{200b}$ for our four WDM and one CDM
      models. The dotted line indicates the CDM main halo density
      profile from Fig.~\ref{DensityProf}, renormalised to pass through
      the locus of radial distribution points at
      250kpc. \emph{Bottom:} the number fraction of subhaloes per logarithmic
      interval in radius, on a linear-log plot. The area under the
      curves is proportional to subhalo number, so this plot shows
      that subhaloes are preferentially found in the outer parts of the
      halo. The black line corresponds to the CDM model, CDM-W7, while
      the blue, green, orange and red lines correspond to the
      $m_{2.3}$, $m_{2.0}$, $m_{1.6}$, and $m_{1.5}$ WDM models
      respectively.}  
\label{SpatialAB} 
\end{figure} 

\begin{figure}
    \includegraphics[angle=-90,scale=0.33]{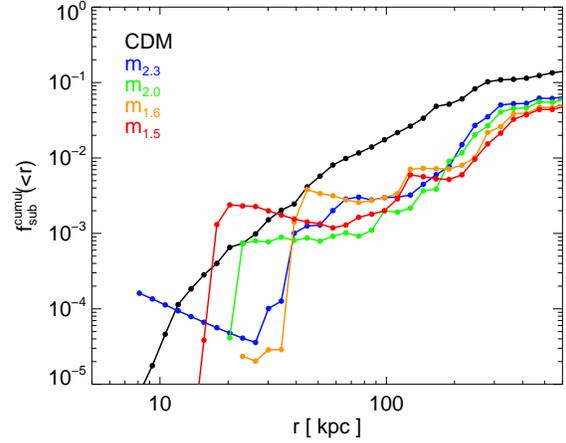}
    \caption{Cumulative mass fraction in substructures as a function
      of radius. The black line corresponds to the CDM model, CDM-W7,
      while the blue, green, orange and red lines correspond to the
      $m_{2.3}$, $m_{2.0}$, $m_{1.6}$, and $m_{1.5}$ WDM models
      respectively.}  
    \label{SubFrac} 
 \end{figure} 

The number density of subhaloes of mass $M_{\rm{sub}}>10^{8}\Msun$ as a
function of radius, normalized to the mean number density within
$r_{200b}$, is shown in the top panel of Fig.~\ref{SpatialAB}. The
bottom panel shows the cumulative number fraction of subhaloes per
logarithmic radial interval. The number density profiles of subhaloes
in the different WDM models are very similar to one another and to the
CDM case. This uniformity is suprising since, as we shall see below,
the central densities of WDM subhaloes decrease with decreasing WDM
particle mass, making them increasingly vulnerable to tidal
disruption. This result is reminiscent of that found by
\citet{Springel08b} that the number density profiles of Aquarius
subhaloes are essentially independent of subhalo mass.  It may be that
better statistics might reveal differences in the radial distribution
of WDM subhaloes.  

The subhalo number density profiles are shallower than that of the
halo dark matter.  \citet{Springel08b} found that the subhalo profiles
are well described by an Einasto form (see Eqn.~\ref{Ein} below), with $r_{-2} = 199 {\rm kpc} =
0.81 r_{200}$ and $\alpha_{\rmn{ein}} = 0.678$.  The lower panel of
Fig.~\ref{SpatialAB} shows that, as was the case for CDM, subhaloes lie
preferentially in the outer parts of the halo, between 100~kpc and the
virial radius, even though the number density is highest in the
central regions.

The cumulative mass fraction in subhaloes as a function of radius is
depicted in Fig.~\ref{SubFrac}. As expected from the mass functions of
Fig.~\ref{CmlVmaxAb}, the subhalo mass fractions in the WDM models are
lower than for CDM. At $r_{200b}$, the mass fractions in WDM subhaloes
are approximately 5 percent, less than half the value in the CDM case. There
is a small, but systematic decrease in the mass fraction with
decreasing WDM particle mass.

 \subsection{The internal structure of WDM subhaloes}
\label{ResSS2} 

We now consider the internal structure of WDM haloes, particularly
their radial density profiles. We begin by performing a convergence
test of the profiles.  

\subsubsection{Convergence of the density profiles}

\begin{figure*}
    \includegraphics[scale=0.80]{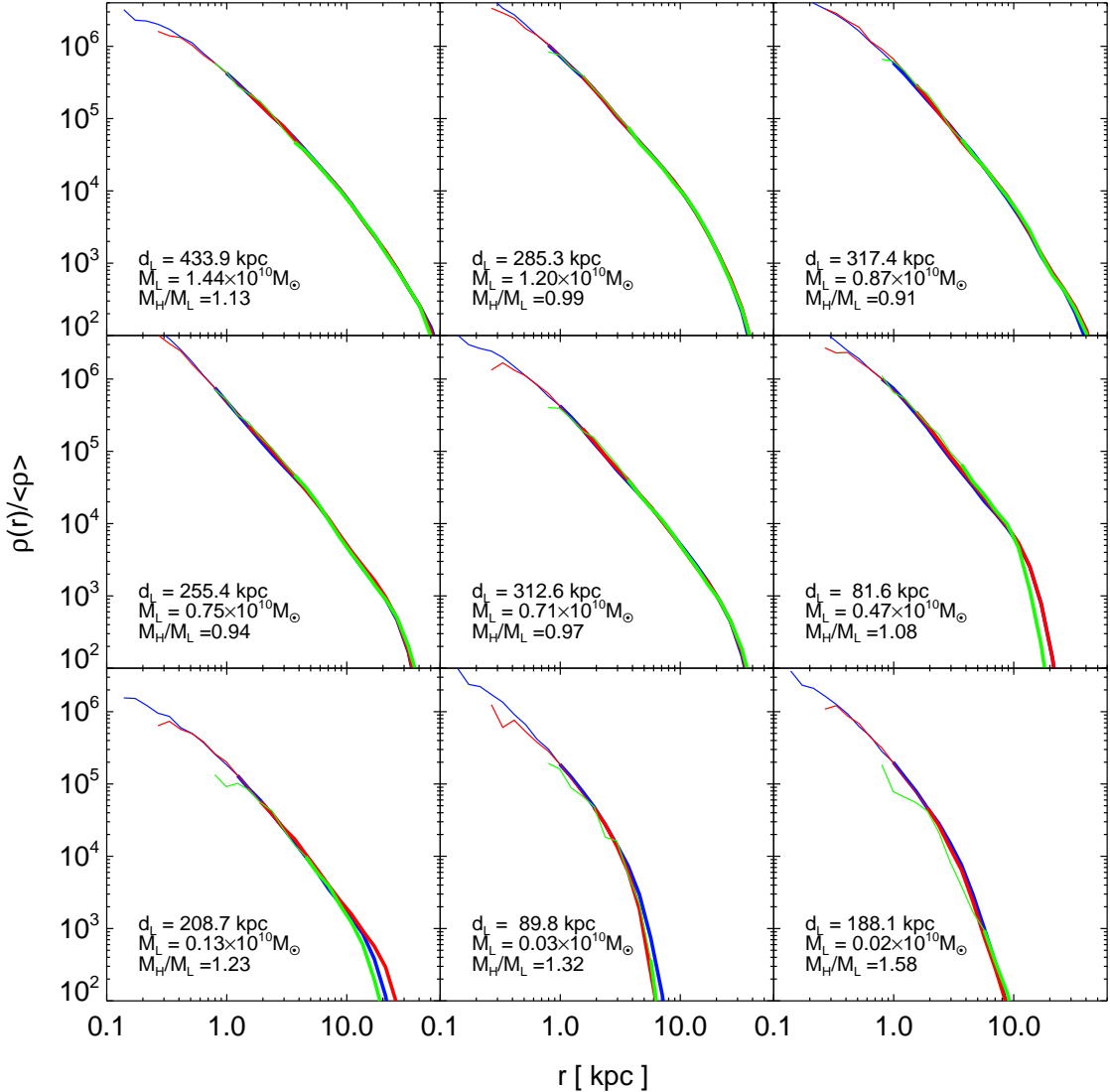}
    \caption{Spherically averaged radial density profiles for subhaloes
      matched between the high (level~2), intermediate (level~3), and
      low (level~4) resolution versions of the $m_{1.5}$ simulation.
      Blue corresponds to high, red to intermediate, and green to low
      resolution. The density profiles are shown by thick lines down
      to the smallest radius at which they satisfy the convergence
      criterion of \citet{Power03}, and are continued by thin lines
      down to a radius equal to twice the softening length. In the
      legend, $d_{L}$ is the distance of the low resolution subhalo
      from the main halo centre, $M_{L}$ is the subhalo mass, and
      $M_{L}/M_{H}$ is the ratio between the masses of the low and
      high resolution counterparts.}  
 \label{ConvDens} 
 \end{figure*}

\citet{Springel08b} carried out a careful study of the
convergence properties of the CDM Aquarius haloes upon which our set of
WDM halo simulations is patterned. Here we carry out an analogous
study of the WDM subhaloes. We focus on the most extreme case,
$m_{1.5}$, since this differs most from CDM. Fig.~\ref{ConvDens} shows
the density profiles of the nine most massive subhaloes lying within
500~kpc in the $m_{1.5}$ simulation at three different resolutions
(levels~2, 3 and 4). For the subhaloes of mass $>1\times10^{9}\Msun$,
we find that the three realizations agree extremely well at all radii
satisfying the convergence criterion of \cite{Power03}. For those of
lower mass, the low resolution (level~4) examples have fewer that
10000 particles and although this limits the range where the
convergence test is applicable, the convergence is still very good.

\begin{figure*}
    \includegraphics[scale=0.80]{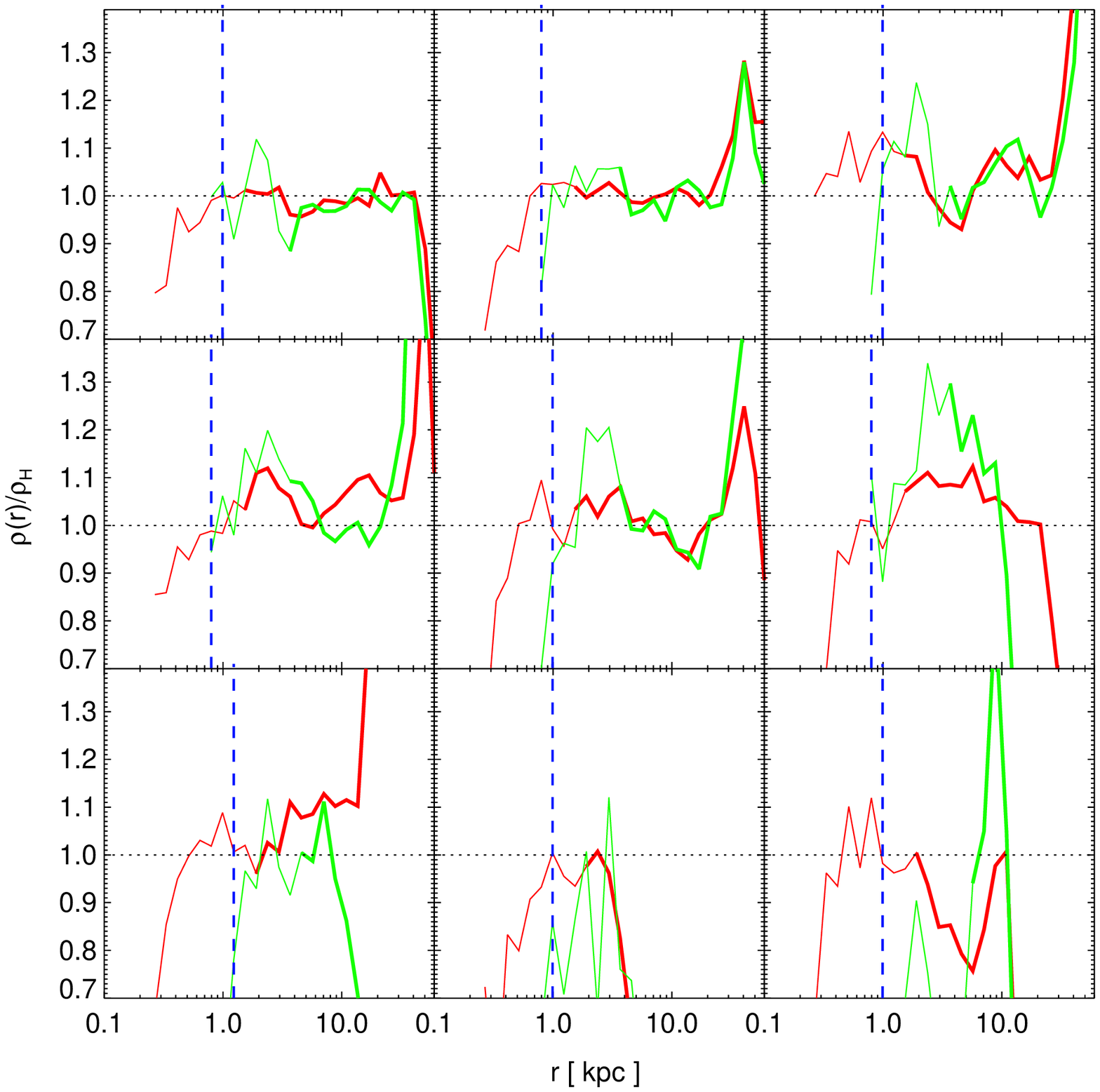}
    \caption{Ratio of the intermediate (level~3; red) and low
      (level~4; green) resolution density profiles of the $m_{1.5}$
      subhaloes shown in Fig.~\ref{ConvDens} to the density profile
      of their high resolution (level~2) counterparts. The blue dashed
      line indicates the convergence radius for the high resolution
      subhaloes.}  
\label{ConvRatio} 
\end{figure*}

  To emphasize the differences between subhaloes simulated at different
  resolution, we plot, in Fig.~\ref{ConvRatio}, the ratios of the
  intermediate and low resolution density profiles to that of their
  high resolution counterparts. At the smallest radius that satisfies
  the \cite{Power03} criterion, the level~3 simulations are converged
  to better than 10 percent; in most cases the same is true of the level~4
  simulations. There are large excursions, however, in the outer
  parts, beyond $\sim 10$~kpc. These are particularly noticeable for
  those subhaloes that are closer than 100~kpc from the main halo
  centre, and reflect the slightly different positions within the main
  halo of each of the matched subhaloes.  

  We can determine the mass range where the density profiles are
  converged by considering the ratio of circular velocities at the
  convergence radius of \cite{Power03} between matched subhaloes at
  different resolution.  Demanding that deviations from the level-2
  simulation should not exceed 10 percent, we find that the structure of
  level-3 subhaloes is well converged for subhalo masses $>10^{8}\Msun$
  whereas for level-4 subhaloes convergence is only achieved for masses
  $>10^{9}\Msun$.

\subsubsection{The density profiles of subhaloes}

We now consider the spherically averaged radial density profiles of
subhaloes in all four different WDM models. For the CDM case
\citet{Springel08b} found that the profiles of subhaloes are well fitted
by either an NFW \citep{NFW_96, NFW_97} or an Einasto
\citep{Einasto65, Navarro04} functional form. The NFW profile is given
by:

\begin{equation} \rho(r) = \frac{\delta_c \,\rho_{\rm
      crit}}{({r}/{r_s})({r}/{r_s}+1)^2}, 
\end{equation} 

\noindent where $\delta_c$ is a characteristic overdensity (usually
expressed in units of the critical density) and $r_s$ is a spatial
scale that marks the transition between the asymptotic slopes of $-1$
and $-3$.  The Einasto profile is given by:

\begin{equation} \rho(r) = \rho_{-2}\,
  \exp \left(- \frac{2}{\alpha_{\rm{ein}}}\left[
      \left(\frac{r}{r_{-2}}\right)^{\alpha_{\rm{ein}}}-1\right]
  \right), 
\label{Ein}
\end{equation} 

\noindent where $r_{-2}$ is the scale (analogous to $r_s$) where the
profile attains a slope of $-2$, $\rho_{-2}$ is the density at
$r_{-2}$ and $\alpha_{\rm{ein}}$ is a shape parameter. Springel et al.
find that Einasto fits (which have an additional free parameter) are
marginally better than NFW fits for CDM subhaloes even when
$\alpha_{\rm{ein}}$ is fixed to a constant.  

\begin{figure}
    \includegraphics[angle=-90,scale=0.33]{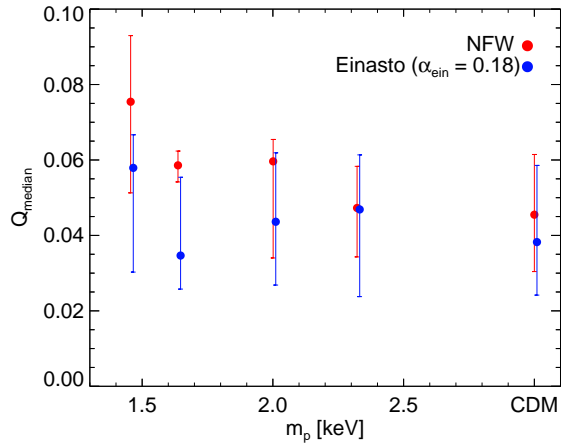}
    \caption{Median value of the goodness of fit statistic, $Q$, for
      Einasto (blue dots) and NFW (red dots) fits to all subhaloes
      of $M_{\rmn{sub}}>10^{9}\Msun$, as a function of the WDM
      particle mass, $m_{\rmn{WDM}}$. In the Einasto fits, we have fixed
      $\alpha_{\rmn{ein}}=0.18$. The error bars indicate the upper and
      lower quartiles of the distribution. The Einasto
      data points are slightly offset in $m_{\rmn{WDM}}$ for clarity.}
    \label{QPlot} 
\end{figure}

Following \citet{Springel08b} we define a goodness of fit statistic
for the functional fits to the subhalo profiles as:

\begin{equation} Q^{2} =
  \frac{1}{N_{\rmn{bins}}}\sum_{i}[\ln\rho_{i} -
    \ln\rho^{\rmn{model}}(r_{i})]^{2}, 
\end{equation} 

\noindent where $\rho_{i}$ is the density measured at radius
$r_{i}$, and $\rho^{model}$ is the model density evaluated at that
same radius.  In Fig.~\ref{QPlot} we show how well our subhaloes can
be fit by NFW and Einasto profiles, in the latter case with fixed
shape parameter ($\alpha_{\rmn{ein}}=0.18$, following
\citealt{Springel08b}), by plotting the median value of $Q$ for each
of the different models as a function of the thermal equivalent WDM
particle mass. As for CDM, we find that the Einasto profile is a
marginally better fit to WDM subhaloes than the NFW profile. There is
little variation in the quality of the Einasto fits for the
different values of the particle mass, but the NFW fits seem to
become slightly worse with increasing mass.

\begin{figure*}
    \includegraphics[angle=-90,scale=0.80]{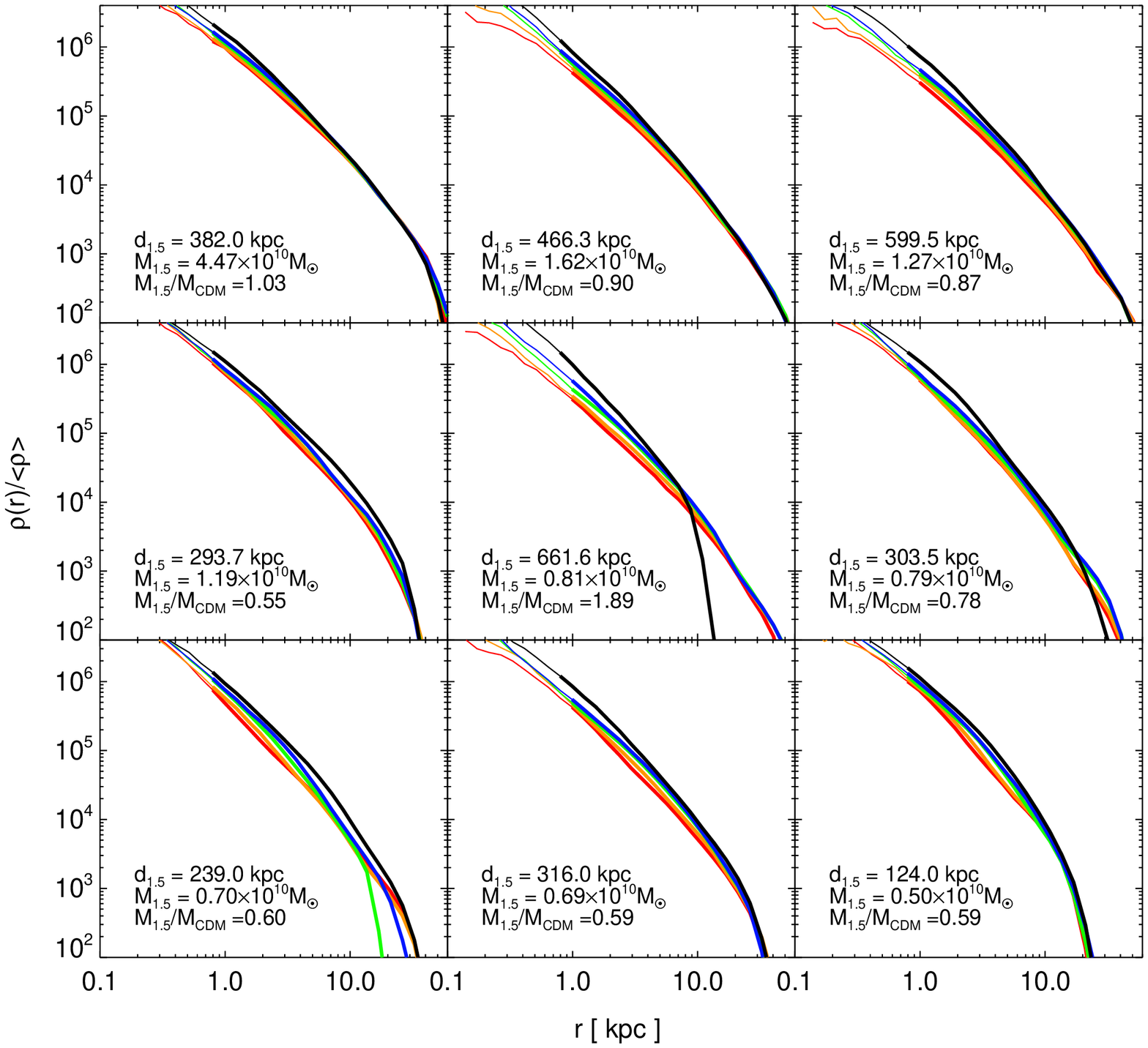}
    \caption{Spherically averaged radial density profiles of subhaloes
      in simulations of different WDM particle mass. The subhaloes have
      been matched across simulations on the basis of their position
      and mass. However, it should be noted that in some cases the
      matches are uncertain. The different colours correspond to
      different WDM particle masses: red, orange, green and blue to
      1.5, 1.6, 2 and 2.3 keV respectively, while black corresponds to
      the CDM case. In the legend, $d_{1.5}$ is the distance of the
      subhalo from the main halo centre in the $m_{\rm WDM}=1.5$keV,
      $M_{1.5}$ is the mass of the subhalo also in this case, and
      $M_{1.5}/M_{{\rm CDM}}$ is the ratio of this mass to that of the
      CDM counterpart. As in Fig.~\ref{ConvDens} the density profiles
      are shown by thick lines down to the smallest radius at which
      they satisfy the convergence criterion of \citet{Power03}, and
      are continued by thin lines down to a radius equal to twice the
      softening length} 
\label{DensMatched} 
\end{figure*}

\begin{figure*}
    \includegraphics[angle=-90,scale=0.80]{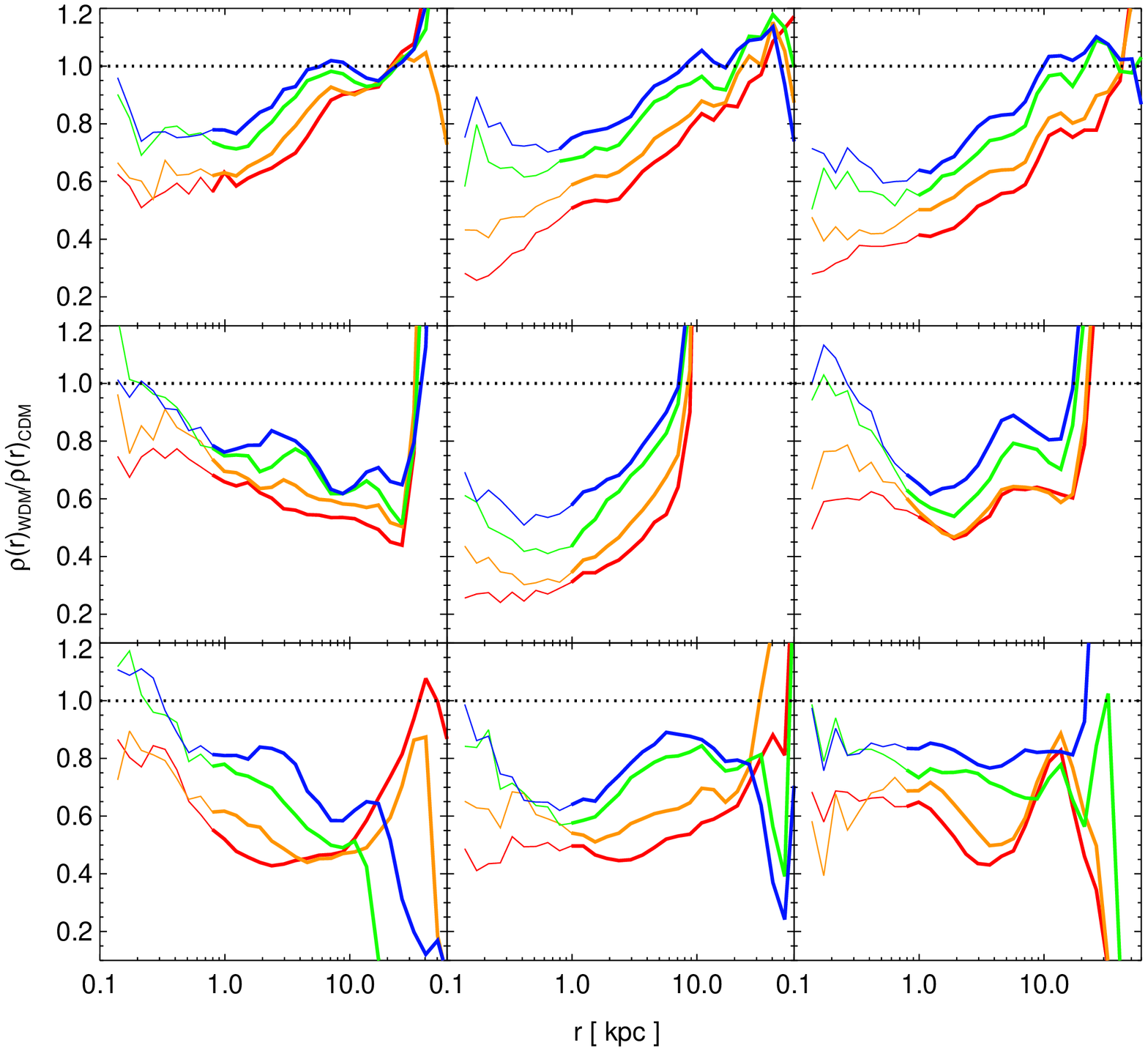}
    \caption{Ratio of the density profiles of matched subhaloes in
      simulations of different WDM particle mass relative to the mass
      of the CDM counterpart. The colours are as in
      Fig.~\ref{DensMatched} as is the use of thick and thin lines.}
    \label{MatchedRatio} 
\end{figure*}

The density profiles of subhaloes vary systematically with the WDM
particle mass. Before performing a statistical comparison, we
illustrate this variation with a few examples of subhaloes that we have
been able to match across simulations with different WDM particle
masses. Such matches are not trivial because the subhaloes have masses
close to the cutoff in the initial power spectrum and thus their
formation histories can vary substantially from one case to another.
In Fig.~\ref{DensMatched} we show nine examples of subhaloes where,
based on their positions and masses, we have been able to identify
likely matches. In Fig.~\ref{MatchedRatio} we show the ratio of the
profiles to that of their CDM counterpart.  

The differences amongst the profiles tend, in most cases, to be larger
at smaller radii. As the WDM particle mass decreases, the subhalo
profiles tend to become shallower. At the innermost converged point,
the density of the subhalo with the smallest value of $m_{\rm WDM}$ is
generally a factor of several smaller than its CDM counterpart. For
example, the $m_{1.5}$~keV subhalo in the central panel of the
Figs.~\ref{DensMatched} and~\ref{MatchedRatio} is a factor of $\sim 3$
less dense at the innermost converged point than its CDM counterpart
and a factor of $\sim 2$ less dense than the subhalo with
$m_{2.3}$keV.

\begin{figure*}
   \includegraphics[scale=0.6, angle=-90]{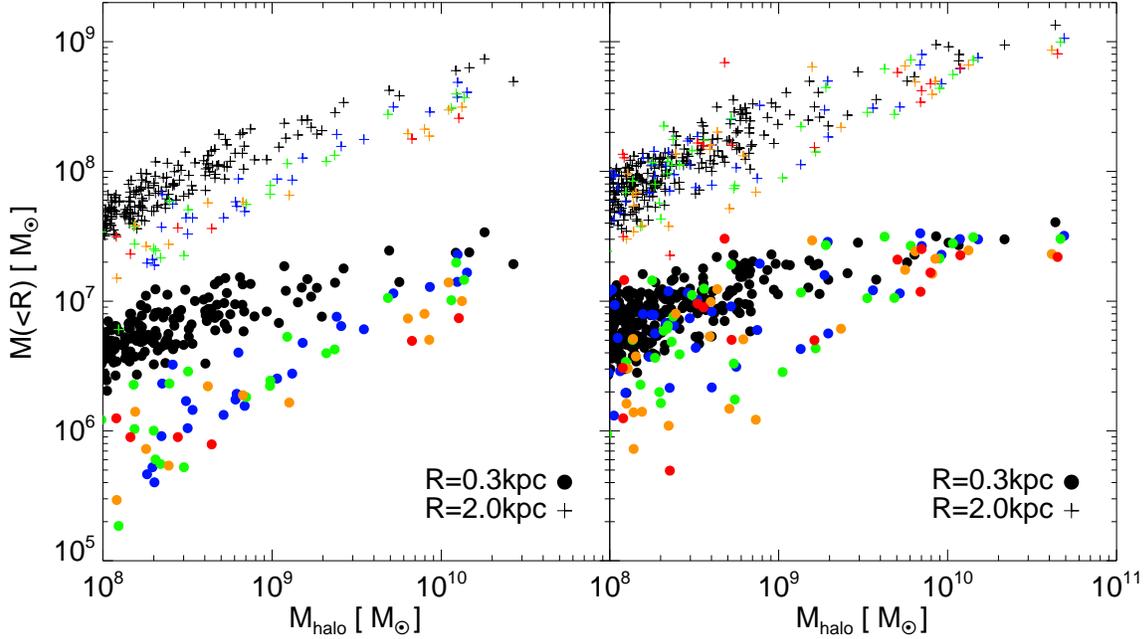}
   \caption{Central masses of field haloes (left) and subhaloes within
     $r_{200b}$ (right), evaluated within radii of 2~kpc (crosses) and
   300pc (circles) as a function of total mass. Different colours
   correspond to different simulations: black for CDM-W7, blue, green,
   orange and red for models $m_{2.3}$, $m_{2.0}$, $m_{1.6}$, and
   $m_{1.5}$ respectively. }
   \label{MEnc}
 \end{figure*}

The trends seen in Figs~\ref{DensMatched} and~\ref{MatchedRatio}
reflect the fact that, for fixed cosmological parameters, haloes of a
given mass form later in WDM models than in CDM \citep{AvilaReese01,
  Lovell12}. We
can quantify the difference by comparing, for example, the central
masses of haloes in our various models.  The masses enclosed within
300~pc and 2~kpc of the centre in field haloes and subhaloes in our
simulations are plotted in Fig.~\ref{MEnc} as a function of halo mass.
For field haloes (left panel) there is a clear separation at both radii
amongst the different models: at fixed mass, the WDM haloes have lower
central masses than their CDM-W7 counterparts and the enclosed mass
decreases with the WDM particle mass.  For (field) haloes of mass less
than $5\times10^{9}\Msun$, the masses enclosed within 300pc are lower
relative to the CDM case by factors of $\sim 4$ and $\sim 3 $ in the
$m_{1.6}$ model and $m_{2.3}$ models respectively. At higher masses
the differences are smaller (by factors of 2 and 3 for the $m_{2.3}$
and $m_{1.6}$ cases respectively), thus the main halo density profiles
varies very little for this range of $m_{\rmn{WDM}}$.  The situation is somewhat
different for subhaloes (right panel), largely because tidal stripping
removes material from the outer regions, leaving the central density
largely unaffected. As a result, after falling into their host halo,
objects move primarily to the left in Fig.~\ref{MEnc} but the change
is comparatively greater for the less concentrated WDM subhaloes than
for the CDM subhaloes. Nevertheless, an offset amongst the WDM subhaloes
and amongst these and the CDM subhaloes remains, particularly at large
masses.

\begin{figure}
    \includegraphics[angle = -90., scale=0.33]{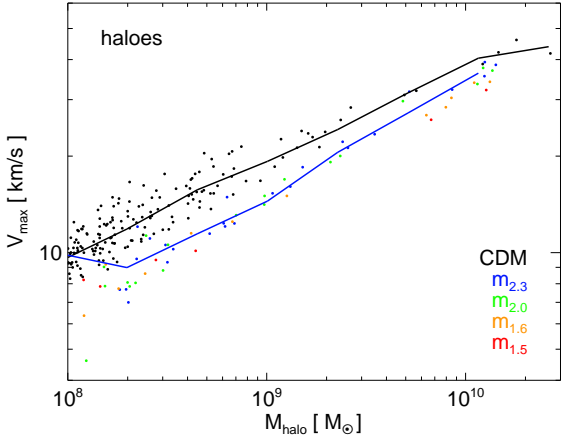}
    \caption{$M_{\rmn{halo}}$ vs. \vmax for field haloes. The black
      dots show the data for the CDM-W7 simulation and the black line
      represents the mean relation. The colour dots show data for the
      WDM simulations: blue, green, orange and red for models
      $m_{2.3}$, $m_{2.0}$, $m_{1.6}$, and $m_{1.5}$ respectively.
      The mean relation is shown only for the $m_{2.3}$ WDM model in
      which the number of subhaloes is largest and thus the least
      noisy.}
   \label{VmaxMsub}
 \end{figure}

Another measure of central mass is provided by the value of \vmax
which we plot as a function of mass for field haloes in
Fig.~\ref{VmaxMsub}. There is a marked difference between the CDM-W7
and the WDM haloes which, at a given mass, have a lower $V_{\rm max}$.
As expected, these differences decrease with increasing halo mass. At
$10^9\Msun$ the mean value of \vmax for the $m_{2.3}$ case is a factor
of 1.33 smaller than for CDM-W7.

\begin{figure}
   \includegraphics[angle=-90,scale=0.33]{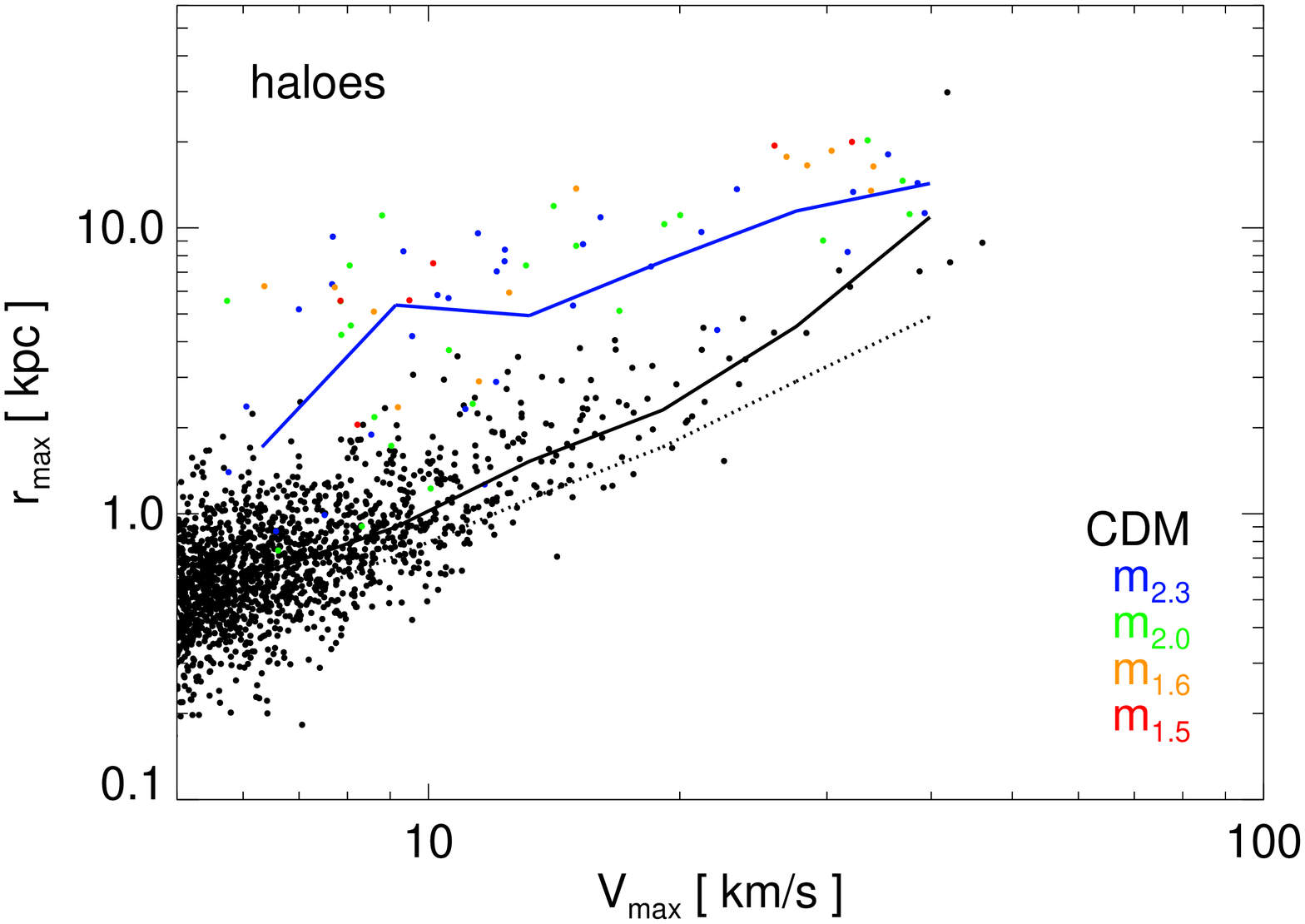}
   \includegraphics[angle=-90,scale=0.33]{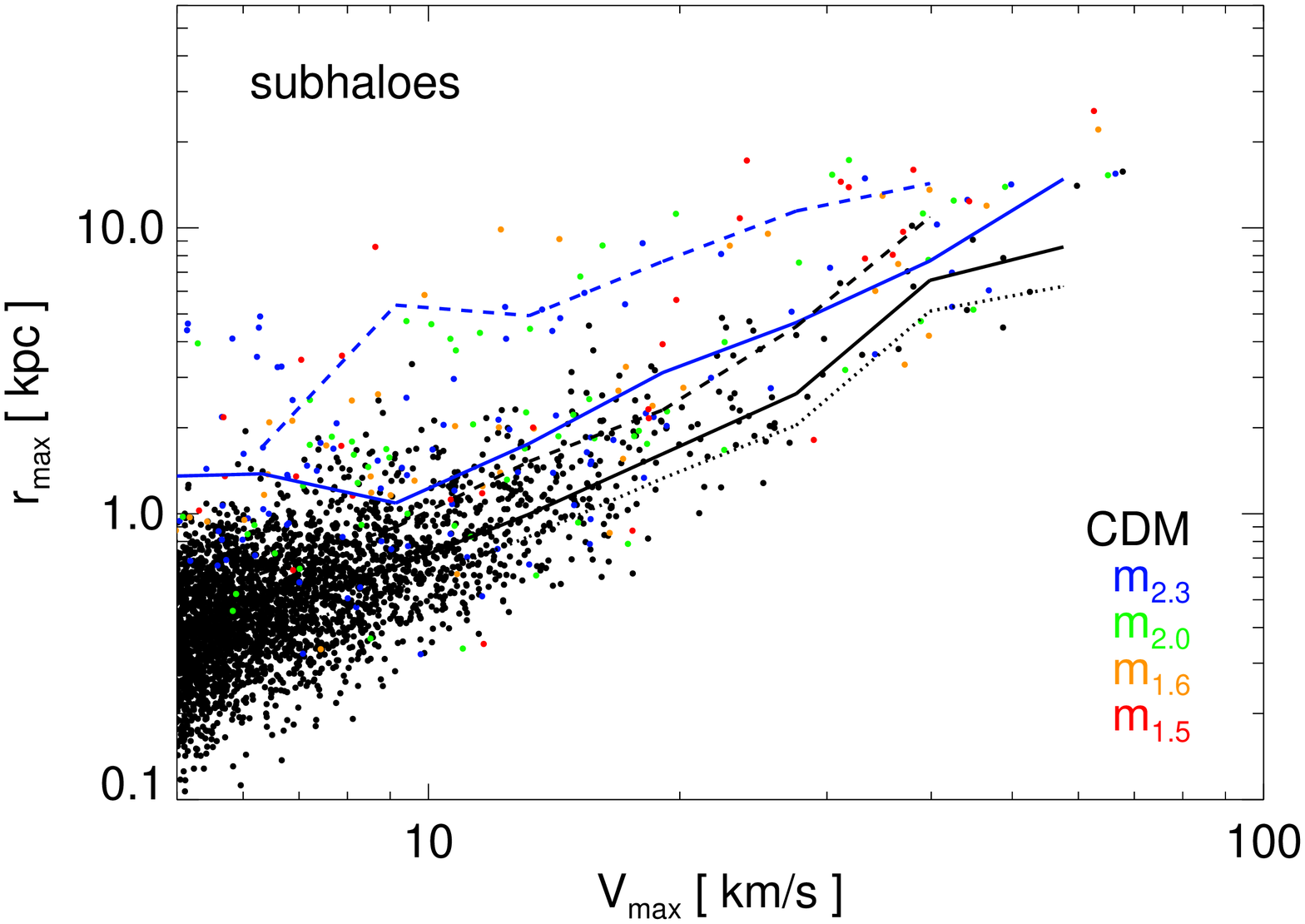}
   \caption{\vmax vs. $r_{\mathrm{max}}$ for independent haloes (top)
     and subhaloes (bottom).  The black dots show the data for the
     CDM-W7 simulation and the black line represents the mean relation
     in the case.  The dotted line corresponds to a $\Lambda$CDM
     simulation using the WMAP1 cosmological parameters.  The colour
     dots show data for the WDM simulations: blue, green, orange and
     red for models $m_{2.3}$, $m_{2.0}$, $m_{1.6}$, and $m_{1.5}$
     respectively.  The mean relation is shown only for the $m_{2.3}$
     WDM model in which the number of subhaloes is largest and thus the
     least noisy.  The solid lines of the top panel are reproduced in
     the bottom panel as dashed lines.}
   \label{VmaxRmax}
 \end{figure}

The differences in the internal structure of haloes in the WDM and CDM
cases can be further quantified by comparing the relation between
\vmax and $r_{\rmn{max}}$, the radius at which \vmax is attained. We
plot these relations separately for independent haloes and subhaloes in
Fig.~\ref{VmaxRmax}.  Tidal stripping of CDM subhaloes causes their value
of \vmax to drop less rapidly than their value of $r_{\rmn{max}}$,
leading to an increase in the concentration of the subhalo
\citep{Penarrubia08,Springel08b}. As may be seen by comparing the top
and bottom panels of Fig.~\ref{VmaxRmax}, the values of
$r_{\rmn{max}}$ for CDM subhaloes at fixed \vmax are typically
70 percent of the values for field haloes
\footnote{This number depends on the choice of cosmological
  parameters. For the Aquarius simulations (which assumed WMAP1
  cosmological parameters), this number decreases to 62 percent
  \citep{Springel08b}, as can be seen by comparing the dotted lines in
  the two panels of Fig.~\ref{VmaxMsub}.  This difference is driven
  primarily by the higher value of $\sigma_{8}$ in the WMAP1 cosmology
  which causes haloes of a given mass to collapse earlier and thus be
  more concentrated than their WMAP7 counterparts.}.  Since WDM
subhaloes are less concentrated than their CDM counterparts to begin
with, they are more susceptible to stripping once they become subhaloes
\citep[see also][]{Knebe02}. Thus, at fixed $V_{\rm max}$, the values
of $r_{\rmn{max}}$ in the $m_{2.3}$ case are now typically only 40 percent
of the values for field haloes.
Even so, since the typical values of $r_{\rmn{max}}$ for subhaloes with
$V_{\rm max}>10$~km/s are greater than $1\rmn{kpc}$ (even in the
models with the smallest WDM particle mass), the majority of any dSphs
residing in subhaloes like these would not show clear signs of tidal
disruption.

\subsection{The abundance of the most massive subhaloes}

\cite{BoylanKolchin11,BoylanKolchin12} showed that the most massive
subhaloes in the Aquarius halo simulations are much too massive and
concentrated to host the brightest dSph satellites of the Milky Way.
\cite{Parry2012} reached the same conclusion using gasdynamic
simulations of the Aquarius haloes. This discrepancy was called the
`too big to fail problem' by Boylan-Kolchin et al. Subsequently
\cite{Wang12} showed that the extent of the discrepancy depends
strongly on the mass of the Galactic halo and all but disappears if
the Milky Way's halo has a mass of $1\times 10^{12}\Msun$,. Alternatively,
\citet{Lovell12} showed the the problem is naturally solved in a WDM
model even if the mass of the Galactic halo is $2\times 10^{12}\Msun$.
Their WDM model, chosen to have a particle mass only just compatible
with the Lyman-$\alpha$ constraints of \cite{Boyarsky2009c,
  Boyarsky09b} (but not with the more recent constraint quoted by
\citealt{Viel_13}) is the $m_{1.5}$ model of the current study.

The Milky Way contains three satellites, the LMC, SMC and Sagittarius,
that are brighter than the brightest dSph, Fornax. The `too big to
fail problem' consists of having substantially more than three
massive subhaloes within 300~kpc in the simulations whose properties
are incompatible with the measured kinematics of the nine brightest
dSphs, specifically with the measured masses within their half-light
radii (where masses can be robustly measured from the data;
\citealt{Walker09, Wolf10}).  In our WDM simulations we thus count the
number of subhaloes within 300~kpc of the main halo centre that have
circular velocity profiles of amplitude greater than the measured
half-light circular velocities of the 9 brightest dSphs plus their
$3\sigma$ errors \citep{Walker09, Wolf10, Lovell12}. We find 1, 1, 3
and 4 subhaloes in the $m_{1.5}$, $m_{1.6}$, $m_{2.0}$ and $m_{2.3}$
WDM models respectively and 6 in CDM-W7. Thus, all our WDM simulations
are free of the `too big to fail problem' even in a $2\times
10^{12}\Msun$ Galactic halo. Note that if we knew the mass of the
Milky Way halo precisely, this argument could, in principle, be used
to set an {\em upper limit} on the (thermal) WDM particle mass.

 \section{Discussion and conclusions}
 \label{Conc}
 Although the existence of dark matter was inferred in the 1930s, its
 identity remains one of the most fundamental unsolved questions in
 physics. The evidence points towards dark matter being made of as yet
 undiscovered elementary particles. Over the past thirty years
 attention has focused on cold dark matter
 \citep{Peebles_82,Davis85,Bardeen1986} but this is not the only
 possibility. For example, the lightest sterile neutrino in the
 $\nu$MSM model \citep{Asaka_05} would behave as warm dark matter,
 generating very similar structures to CDM on scales larger than
 bright galaxies but very different structures on smaller scales
 \citep{Lovell12,Maccio12,Schneider2012}.

 In this study we have carried out a series of high resolution N-body
 simulations of galactic haloes in universes dominated by WDM, taking
 as the starting point one of the haloes from the Aquarius project of
 simulations of CDM galactic haloes carried out by the Virgo Consortium
 \citep[``Aq-A'' in][]{Springel08b}. As a prelude we
 resimulated this CDM halo replacing the cosmological parameters from
 the WMAP year-1 values assumed by Springel et al. to the WMAP year-7
 values \citep{wmap11}. For CDM this change has the effect of
 lowering the central densities of galactic subhaloes, alleviating (but
 not eliminating) the tension between the structure of CDM subhaloes
 orbiting in haloes of mass $\sim 2\times 10^{12}\Msun$ and the
 kinematical data for Milky Way satellites
 \citep{BoylanKolchin12,Wang12}. We then performed a series of
 simulations of WDM haloes, using as initial conditions the same
 fluctuation phases and linear power spectrum of Aq-A, suitably
 truncated to represent WDM with (thermal equivalent) particle masses
 in the range $1.5$~keV to $2.3$~keV. Our main simulations correspond to
 level-2 resolution in the notation of \cite{Springel08b}, but we
 also ran simulations at lower resolution to establish
 convergence. 

 N-body simulations with a resolved cutoff in the initial power spectrum
 undergo artificial fragmentation in filaments \citep{Bode01,Wang07}.
 The resulting spurious structures need to be identified before the
 simulations can be analyzed. This is best done in the initial
 conditions: we found that the spurious fragments evolve from
 disc-like structures that are much flatter than the progenitors of
 genuine haloes. The sphericity of structures in the initial conditions
 therefore provides a robust flag for spurious objects which we
 supplement with a mass cut, $M_{\rmn{min}}$, derived from the
 limiting mass for genuine haloes, $M_{\rmn{lim}}$, inferred by
 \citet{Wang07} from simulations of hot dark matter models. We find
 that a cut of $M_{\rmn{min}}=\kappa M_{\rmn{lim}}$, with $\kappa=0.5$, 
 captures the results from a comparison of matched haloes in
 simulations of different resolution.  The combined sphericity and
 mass cut criteria result in clean catalogues of genuine haloes and
 subhaloes.

 The spherically averaged density profile of the main halo is
 virtually indistinguishable in the CDM and all our WDM simulations
 but there are large differences in the abundance and structure of
 their subhaloes.  For WDM, the subhalo mass functions begin to diverge
 from the CDM case at masses between $\sim 2\times 10^{9}\Msun$ for
 the $m_{2.3}$ (least extreme) and $\sim 7\times 10^{9}\Msun$
 for the $m_{1.5}$ (most extreme) models. The cumulative mass 
 functions are well fit by fitting functions given in 
 \S\ref{MassVmax}: they become essentially flat for subhaloes
 masses below $\sim 7\times 10^{9}\Msun$. The mass fraction in
 substructures within $r_{200b}$ is lower than in the CDM case by
 factors between 2.4 (for $m_{1.5}$) and 2 ($m_{2.3}$). The radial
 distributions of subhaloes are very similar to the CDM case. 

 WDM haloes and subhaloes are cuspy (except in the very inner regions -
 see \cite{Maccio12} and \cite{Shao13}) and are well fit by NFW
 profiles, and even better by Einasto profiles. However, the central
 density of WDM haloes depends on the WDM particle mass: in those cases
 where it is possible to identify the same subhalo in CDM and
 different WDM simulations, the density profiles have systematically
 shallower slopes in the latter which become flatter for smaller
 particle masses.  This change of slope is reflected in the main halo
 mass, $M_{\rmn {host}}-M_{\rmn{sub}}$, $M_{\rmn{sub}}-$\vmax, and
 \vmax$-r_{\rmn{max}}$ relations, such that, for a given mass,
 subhaloes in warmer dark matter models have progressively lower
 central densities, lower values of \vmax and higher values of
 $r_{\rmn{max}}$ relative to CDM subhaloes. These differences affect
 the evolution of subhaloes once they fall into the main halo since
 less concentrated haloes are more easily stripped.

 Both the abundance and the structure of WDM subhaloes can be compared
 to observational data. The requirement that the models should produce
 at least as many subhaloes as there are observed satellites in the
 Milky Way sets a lower limit to the WDM particle mass. This is a very
 conservative limit since feedback processes, arising from the
 reionization of gas in the early universe and supernova energy, would
 prevent the formation of galaxies in small mass haloes just as they do
 in CDM models \citep[e.g.][]{Benson_02}. However, the number of
 subhaloes above a given mass or \vmax depends, of course, on the host
 halo mass \citep{Gao_04,Wang12}. For the case we have considered, in
 which $M_{\rmn host} \sim 10^{12}\Msun$, we find that the WDM
 particle mass must be greater than 1.5~keV or 1.6~keV depending on
 whether we simply consider the observed number of satellites or apply
 a correction for the limited area surveyed by the SDSS. This limit is
 less stringent than that limit of 3.3~keV (2$\sigma$) inferred by
 \cite{Viel_13} from the clumpiness of the Lyman-$\alpha$ forest of a
 sample of quasars at redshift $z>4$, although the two results are not
 directly comparable because \cite{Viel_13} use a slightly different
 transfer function.  In principle it might also be
 possible to set an upper limit on the WDM particle mass by comparing
 the subhalo central densities with those inferred for the brightest
 satellites of galaxies like the Milky Way.  Current kinematical data
 are insufficient for this test but they are compatible with the
 properties of the most massive subhaloes in the four WDM models we
 have considered none of which suffers from the `too big to fail'
 problem highlighted by \cite{BoylanKolchin12}.

 WDM remains a viable alternative to CDM, along with other
 possibilities such as self-interacting dark matter
 \citep{Vogelsberger12} and cold-plus-warm mixtures
 \citep{Anderhalden13}.  Further theoretical work, including
 simulations and semi-analytical calculations \citep{Benson13,Kennedy13} combined with better data for dwarf galaxies offer
 the prospect of ruling out or validating these models.

 \section*{Acknowledgements}

 ML acknowledges an STFC studentship (STFC grant reference
 ST/F007299/1). CSF acknowledges ERC Advanced Investigator grant
 COSMIWAY.  We would like to thank the anonymous referee for a
   careful reading of the text. The simulations used in this paper
 were carried out on the Cosmology Machine supercomputer at the
 Institute for Computational Cosmology, Durham.  LG acknowledges
 support from the One hundred talents Program of the Chinese Academy
 of Science (CAS), the National Basic Research Program of China
 (program 973, under grant No. 2009CB24901), {\small NSFC} grant No.
 10973018, and an STFC Advanced Fellowship.  This work used the
   DiRAC Data Centric system at Durham University, operated by the
   Institute for Computational Cosmology on behalf of the STFC DiRAC
   HPC Facility (www.dirac.ac.uk). This equipment was funded by BIS
   National E-infrastructure capital grant ST/K00042X/1, STFC capital
   grant ST/H008519/1, and STFC DiRAC Operations grant ST/K003267/1
   and Durham University. DiRAC is part of the National
   E-Infrastructure.    This work was supported in part by an STFC
 rolling grant to the ICC. The data featured in the figures in this
 paper are available on request from ML.

  \bsp

  \bibliographystyle{mn2e}
 \bibliography{bibtex}

   \label{lastpage}

  \appendix
  \section{Convergence study}
  \label{ConStudy}

  For several dwarf spheroidal satellites of the Milky Way it is
  possible to measure the circular velocity at the radius encompassing
  half the light in a relatively model-independent way
  \cite{Walker09,Wolf10}. The smallest measured value is
  5.7~\kms for Leo IV. The circular velocity at the
  half-light radius is a lower bound on $V_{\rmn{max}}$. Therefore, to
  compare with Milky Way data, we need the number of subhaloes in the
  simulations with \vmax greater than $5.7$\kms. It is important to
  check that the simulations resolve all these subhaloes. 

  We have performed a convergence study using the level 4, level 3,
  and level 2 simulations for two of the WDM models. For the $m_{2.3}$
  model, the subhalo \vmax function at level 4 deviates by 10 percent from
  that in the corresponding level 2 simulation at a value of
  \vmax$=11$\kms; the level 3 subhalo \vmax function deviates by the
  same amount at a value of \vmax$=6$\kms. The particle masses in the
  level~4 and level~3 simulations differ by a factor of 8. If we write
  $(m_{4}/m_{3})^{n}=$\vmax$(4)/$\vmax$(3)$ (where the numbers denote
  the resolution level) we find $n=0.29$. The high resolution,
  level~2, simulation has a particle mass 3.6 times smaller than that
  of level 3. Therefore we expect this simulation to be complete to
  10 percent at \vmax$=4.2$\kms. A similar analysis for the $m_{1.5}$
  simulation shows that this is already complete at level 3 for
  \vmax=$5.7$\kms.

  We have checked the validity of this approach by analysing the
  original Aquarius Aq-A2 and Aq-A1 simulations. The Aq-A1 simulation
  has a particle mass of $1.7\times 10^{3}\Msun$, a factor of $\sim 8$
  smaller than the level 2 simulations. We find that at
  \vmax=$5.7$\kms\ the Aq-A2 subhalo \vmax function deviates by 8 percent
  from the Aq-A1 result. The suppression of small subhaloes in WDM
  models should result in better subhalo completeness in this case
  compared to CDM in this mass range (c.f. convergence between
  levels~3 and~2 for $m_{1.5}$). We therefore conclude that we have
  lost no more than 8 percent of the `true' number of subhaloes in the
  $m_{2.3}$ simulation and even fewer in the warmer models. 

\end{document}